\documentclass[preprint,12pt,authoryear]{elsarticle}

\usepackage[T1]{fontenc}
\usepackage[utf8]{inputenc}
\usepackage{amsmath,amssymb}
\usepackage{booktabs}
\usepackage{graphicx}
\usepackage{tabularx}
\usepackage{microtype}
\usepackage{lineno}
\usepackage{url}

\biboptions{authoryear}

\begin{document}

\begin{frontmatter}

\title{Longitudinal fMRI reveals multi-network reconfiguration and a recurrent frontoparietal shift across rest and task states during adult aging}

\author[1,2,3]{Tingshan Liu}
\author[2,3]{Wanxin Dong}
\author[2,3]{Hairong Zheng}
\author[2,3]{Dong Liang}
\author[2,3]{Zhanli Hu}

\author[1]{Guanxun Cheng\corref{cor1}}
\ead{chengguanxun@126.com}

\author[2,3]{Na Zhang\corref{cor1}}
\ead{na.zhang@siat.ac.cn}

\cortext[cor1]{Corresponding authors}

\affiliation[1]{
    organization={Department of Medical Imaging, Peking University Shenzhen Hospital},
    city={Shenzhen},
    country={China}
}

\affiliation[2]{
    organization={Paul C. Lauterbur Research Center for Biomedical Imaging, Shenzhen Institutes of Advanced Technology, Chinese Academy of Sciences},
    city={Shenzhen},
    country={China}
}

\affiliation[3]{
    organization={Key Laboratory of Biomedical Imaging Science and System, State Key Laboratory of Biomedical Imaging Science and System, Chinese Academy of Sciences},
    city={Shenzhen},
    country={China}
}

\begin{abstract}
Functional connectivity changes across adult aging, but most evidence is cross-sectional, and covariance-based connectivity depends on blood oxygenation level dependent (BOLD) fluctuation amplitude. Covariance decline can reflect changing fluctuation amplitude, and its interpretation may depend on whether normalization uses signal amplitude or includes estimated noise. It is also unclear which longitudinal changes recur across brain states. We analyzed 852 adults (baseline age 36--89 years) from the Aging Adult Brain Connectome with repeated functional MRI (fMRI) during rest, inhibitory control, associative memory, and visual--motor processing (2,220 sessions; median follow-up 5.5 years). Baseline age and elapsed follow-up were modeled separately. Ten cortical networks were tested jointly across states, and covariance was normalized either by released signal amplitude or by a candidate total amplitude reconstructed from released signal and noise amplitudes. Native common covariance declined by 7--11\% per 5 years, while signal pair amplitude declined by 9\% on average across states, significantly more than estimated noise pair amplitude. Under signal normalization, the relative balance among networks reconfigured over follow-up and differed across states, and the frontoparietal network shifted positively relative to the other nine networks in all four states, by 5--8\% per 5 years. The estimated common trajectory was near zero under signal normalization but declined by 4--7\% per 5 years under candidate total normalization. The network reconfiguration and the positive frontoparietal-relative direction persisted under both. Within-person coupling between the frontoparietal shift and fluid cognition was not detected. Adult aging thus combines normalization-dependent overall change with a state-dependent redistribution among networks that includes a recurrent frontoparietal shift.
\end{abstract}


\end{frontmatter}


\section{Introduction}

Large-scale functional networks change across the adult lifespan, but the organization of that change remains incompletely understood. Cross-sectional studies have commonly reported weaker within-network organization, stronger interaction among networks, and reduced system segregation with increasing age \citep{Chan2014, Deery2023}. Longitudinal resting-state work has provided within-person evidence for declining segregation in associative systems, including default, frontoparietal control, and salience networks \citep{Malagurski2020}. At the same time, functional architecture is not fixed across brain states. Age differences in network organization can differ between rest and task, and older adults may show less consistent network organization across states \citep{Hughes2020, Varangis2021}. These observations raise a basic longitudinal question: does adult aging produce a systems-level reconfiguration that recurs across distinct functional states, or are longitudinal changes primarily state specific?

A second issue is measurement scale. Covariance-based functional connectivity preserves both temporal coupling and blood oxygenation level dependent (BOLD) fluctuation amplitude, whereas correlation-based measures remove amplitude by standardization. This distinction matters in aging: healthy aging is associated with a broadly distributed reduction in resting-state BOLD amplitude, such that covariance-based connectivity decreases with age while correlation-based connectivity is comparatively preserved \citep{Strain2022}. Standardization itself involves a choice. Correlation normalizes covariance by measured fluctuation amplitudes, which include residual noise. Normalization by estimated signal amplitudes uses a different denominator, and the two can yield different age trajectories when signal and noise components change differently. Relative network comparisons, in turn, remove a multiplicative change shared across all network covariances while remaining sensitive to differences among networks. Whether these components can be separated within individuals over time, and which conclusions depend on how amplitude is normalized, remains unclear.

This distinction also matters for network-specific aging effects. Associative networks, including frontoparietal systems, have repeatedly been implicated in age-related changes in functional segregation and task organization \citep{Malagurski2020,Hughes2020}. A network's relative position can change because its own covariance changes, because other networks change, or both. We therefore evaluated the complete network profile before characterizing individual networks within it.

\begin{figure}
    \centering
    \includegraphics[width=\linewidth]{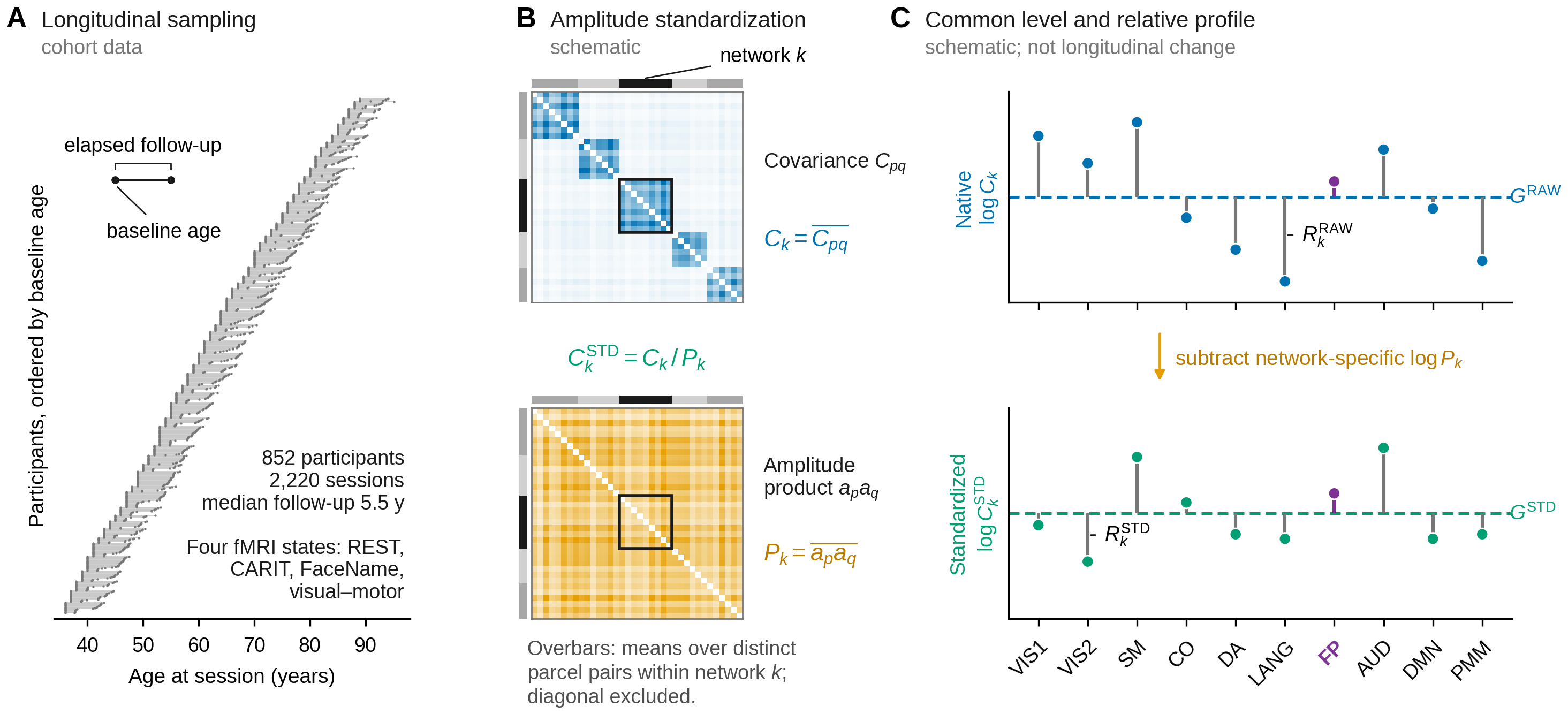}
    \caption{Study design and covariance representations. (A) Longitudinal sampling. Each line runs from a participant's first to last eligible session (intermediate sessions not shown; 852 participants, 2,220 sessions). Where a line starts gives baseline age, and its length gives elapsed follow-up. (B) Schematic matrices for a reduced set of example parcels and networks. Network covariance $C_k$ and pair amplitude $P_k$ are means over distinct parcel pairs within network $k$, excluding the diagonal. Their ratio gives amplitude-standardized covariance. (C) Schematic ten-network profiles. For each representation, $G = (1/10)\sum_k L_k$ is the common level (dashed line) and $R_k = L_k - G$ is the network-relative profile (stems), with $\sum_k R_k = 0$. The two rows show two representations of the same illustrative measurements, not successive visits; differences between the rows do not represent longitudinal change. Sections 3.1 and 3.3 analyze $G$ under native and normalized representations; Sections 3.2–3.3 analyze $R^{\mathrm{STD}}$, including the FP-relative contrast (purple).}
    \label{fig:schematic}
\end{figure}

Here we examined within-person change in network covariance over a median of 5.5 years in 852 adults from the Aging Adult Brain Connectome, with fMRI during rest and three tasks: inhibitory control, associative memory, and visual--motor processing. We separated baseline age from elapsed follow-up, tested ten cortical networks jointly across states, and compared covariance normalized by released signal amplitude with covariance normalized by a candidate total amplitude reconstructed from the released signal and noise estimates. Native covariance declined alongside released signal amplitude, and whether a common decline remained after normalization depended on the amplitude denominator. Beneath this overall change, the relative network profile reconfigured in a state-dependent manner under both normalizations, and frontoparietal covariance rose relative to the other networks in rest and all three tasks. Figure~\ref{fig:schematic} summarizes the design and representations.

\section{Methods}

\subsection{Participants and fMRI states}

We analyzed Aging Adult Brain Connectome (AABC) Release 2 \citep{AABCRelease2}, which contains 3T MRI and phenotypic data from typically aging adults beginning in midlife, with longitudinal visits extending through V4. The HCP-Aging study design and multisite imaging protocol have been described previously \citep{Bookheimer2019,Harms2018}. The present analyses used released imaging-derived measures and longitudinal subject-event information from four acquisition sites: Massachusetts General Hospital (MGH), University of California Los Angeles (UCLA), University of Minnesota (UMinn), and Washington University (WashU). All 852 participants in the analytic cohort had Human Connectome Project in Aging (HCA)-prefixed identifiers. The event-level study label changed from HCA to AABC across eligible visits for 650 participants and was treated as visit-level provenance.

We analyzed four fMRI states: resting state (REST), the CARIT Go/NoGo inhibitory-control task, the FaceName paired-associates memory task, and a visual--motor task involving visual stimulation and button responses. States were modeled jointly so that longitudinal changes shared across states and differences among states could be estimated in one framework.

Participants were required to contribute at least two imaging sessions with finite timing information and positive native within-network covariance for the Default and FP measures. Profile rows were further required to be complete and positive across the ten retained networks. The final cohort contained 852 participants, 2,220 sessions, and 8,727 session--state rows (Table~\ref{tab:cohort}).

\subsection{Imaging-derived measures}

The released fMRI imaging-derived phenotypes were generated from time series processed through the HCP pipelines, including MSMAll registration, spatial-ICA/FIX cleanup, and temporal-ICA cleanup \citep{Glasser2013,Glasser2016,SalimiKhorshidi2014,Glasser2018}. According to the release documentation, structured artifacts related to head motion and physiology were removed before the connectivity products were estimated, and images were intensity normalized \citep{AABCRelease2}. The available release documentation does not specify whether task-evoked responses were regressed out before covariance estimation. Task-state covariance may therefore include shared task-evoked responses. Run-level motion, temporal signal-to-noise ratio, and usable-frame summaries were unavailable in the resources used for this analysis and could not be used as covariates.

We mapped the 360 HCP-MMP1.0 cortical parcels \citep{Glasser2016} to canonical functional networks using the assignment of \citet{Ji2019}. A network was retained if it contained at least two parcels and, in every state, had finite within-network covariance in at least 90\% of eligible parent observations and positive covariance in at least 99\% of finite observations. Ten networks met these criteria: Visual1, Visual2, Somatomotor, Cingulo-Opercular, Dorsal-attention, Language, Frontoparietal (FP), Auditory, Default, and Posterior-Multimodal.

For each participant, visit, and state, the within-network covariance $C_k$ of network $k$ is the mean of the released full-covariance values over distinct parcel pairs within the network. Recomputing it from the released parcel-by-parcel matrices reproduced the network summaries in a diagnostic sample of 12 sessions (Section~S1). The release also provides, for each parcel, a signal amplitude $a_p$ from which unstructured noise has been removed and a separately estimated noise amplitude $n_p$ \citep{AABCRelease2}. We used the released covariance and signal amplitudes instead of the released correlation product because the release documentation recommends covariance and notes that correlation retains spatially nonuniform unstructured noise related to coil sensitivity.

\subsection{Covariance representations}
\label{sec:representations}

Covariance was analyzed on the natural-log scale, $L_{kivs}=\log C_{kivs}$, for network $k$, participant $i$, visit $v$, and state $s$. For each visit--state observation, the common level was the equal-weight mean across the $K=10$ networks and the network-relative profile was the deviation from it:
\begin{equation}
    G_{ivs}=\frac{1}{K}\sum_{k=1}^{K}L_{kivs},
    \qquad
    R_{kivs}=L_{kivs}-G_{ivs}.
\end{equation}
The ten profile values sum to zero within each observation. This centering is a centered log-ratio representation: each value expresses a network's covariance relative to the geometric mean across networks \citep{Aitchison1982}. It removes a multiplicative scale shared by all networks while retaining their relative balance.

\paragraph{Signal normalization} For a network with $m$ available parcels, the signal pair amplitude was the mean product of signal amplitudes over distinct parcel pairs,
\begin{equation}
    P_k=\frac{1}{\binom{m}{2}}\sum_{p<q}a_pa_q
      =\frac{\left(\sum_p a_p\right)^2-\sum_p a_p^2}{m(m-1)},
\end{equation}
requiring amplitude data for at least 75\% of the network's parcels and at least two parcels. The signal-normalized (amplitude-standardized) covariance was $C^{\mathrm{STD}}_k=C_k/P_k$, with $L^{\mathrm{STD}}_k=L_k-\log P_k$. Its common level $G^{\mathrm{STD}}$ and relative profile $R^{\mathrm{STD}}$ were defined by the same centering. Because normalization is applied after averaging within each network, $C^{\mathrm{STD}}$ is a network-level amplitude-standardized covariance instead of a mean Pearson correlation. This was the primary representation for network analyses.

\paragraph{Candidate total normalization} In diagnostic sessions, the diagonals of the released covariance matrices matched the squared signal amplitudes $a_p^2$, and many parcel-pair ratios $C_{pq}/(a_pa_q)$ fell outside $[-1,1]$, consistent with unstructured noise variance being absent from the covariance diagonal (Section~S1). We therefore formed a candidate total parcel amplitude, assuming that signal and unstructured noise are additive and uncorrelated,
\begin{equation}
    t_p=\sqrt{a_p^2+n_p^2},
    \qquad
    P^{\mathrm{TOT}}_k=\frac{1}{\binom{m}{2}}\sum_{p<q}t_pt_q,
    \qquad
    L^{\mathrm{TOT}}_k=L_k-\log P^{\mathrm{TOT}}_k .
\end{equation}
All diagnostic parcel-pair ratios $C_{pq}/(t_pt_q)$ fell within $[-1,1]$, and adding $\mathrm{diag}(n_p^2)$ to the released covariance matrices restored positive semidefiniteness in the three diagnostic REST sessions. Single-run task matrices retained small negative eigenvalues, compatible with reconstruction error affecting rank-deficient task matrices. Because the release does not document the generating estimator, we refer to this representation as candidate total normalization. The two normalizations differ by an attenuation term,
\begin{equation}
    L^{\mathrm{TOT}}_k=L^{\mathrm{STD}}_k+H_k,
    \qquad
    H_k=\log\frac{P_k}{P^{\mathrm{TOT}}_k}\leq 0,
\end{equation}
and this identity carries through the common level, the centered profile, and the FP-relative contrast. A noise pair amplitude $P^{\mathrm{N}}_k$ was defined analogously from $n_p$; in general $P^{\mathrm{TOT}}_k\neq P_k+P^{\mathrm{N}}_k$. The distribution of $P_k/P^{\mathrm{TOT}}_k$ across the cohort is summarized in Table~S2.

\subsection{Longitudinal models and inference}

Baseline age and elapsed follow-up were represented separately so that cross-sectional age differences did not substitute for within-person change. Baseline age was centered at the cohort mean, and both baseline age and elapsed time from the first eligible visit were scaled in 5-year units. The fixed-effects design included baseline age, elapsed time, state, baseline-age$\times$state, elapsed-time$\times$state, sex, study, site--scanner, elapsed-time$\times$site, and elapsed-time$\times$site$\times$state terms. State and site interactions were sum coded. The four state-specific longitudinal slopes were linear combinations of the elapsed-time and elapsed-time$\times$state coefficients, and the equal-state slope was their mean.

Repeated visit--state observations were modeled with a Gaussian working covariance containing participant and participant-by-state components, $V_i=I+\tau_{P}\mathbf{1}\mathbf{1}^{\mathsf T}+\tau_{PS}Z_iZ_i^{\mathsf T}$, where $Z_i$ indicates state membership. Fixed effects were estimated by generalized least squares (GLS), and uncertainty was estimated with a sandwich covariance aggregated at the pedigree level, so that related participants formed one cluster.

The centered profile has nine independent dimensions. We represented it in nine orthonormal Helmert coordinates, which form an isometric log-ratio basis \citep{Egozcue2003}, and fitted these jointly with the same design and an unrestricted residual covariance across coordinates. Coefficients and pedigree-level influence functions were then transformed back to network contrasts.

Joint hypotheses were tested with Wald $\chi^2$ statistics using the pedigree-robust covariance. Unless stated otherwise, intervals for state-specific slopes are Bonferroni simultaneous 95\% intervals across the four states, and tests are two-sided with $\alpha=.05$.

\subsection{Common level and amplitude components}

The native common level $G^{\mathrm{RAW}}$, the common signal pair-amplitude component $G^{A}=K^{-1}\sum_k\log P_k$, and the signal-normalized common level $G^{\mathrm{STD}}=G^{\mathrm{RAW}}-G^{A}$ were each first fitted with their own estimated working covariance. Holding the working covariance of the native model fixed gives the exact coefficient identity $\boldsymbol\beta_{\mathrm{RAW}}=\boldsymbol\beta_{A}+\boldsymbol\beta_{\mathrm{STD}}$. Under this decomposition, the fraction of each native state slope remaining after normalization was $q_s=|\beta^{\mathrm{STD}}_s/\beta^{\mathrm{RAW}}_s|$. A prespecified descriptive criterion required the simultaneous 95\% upper bound of $\max_s q_s$, obtained from 4,999 Gaussian pedigree-cluster multiplier draws, to fall below 0.50.

To compare amplitude components on a common footing, the common log signal, candidate total, and noise pair amplitudes, together with the attenuation term, were fitted with identical GLS weights (the working covariance of the signal-normalized common model). Intervals for their state and equal-state slopes were Bonferroni simultaneous across the resulting family of 20 slopes, and the three amplitude four-state tests were Holm adjusted. Signal and noise trajectories were compared directly with paired contrasts of their slopes (Bonferroni across the four states and the equal-state mean).

\subsection{Network profile and FP characterization}

From the joint profile model, the primary omnibus test evaluated whether equal-state longitudinal slopes differed among the ten networks (time$\times$network; 9 df), and a second test evaluated whether those differences varied across states (time$\times$network$\times$state; 27 df). We estimated all 40 network--state slopes with Bonferroni simultaneous intervals across the 40. Profile contraction was summarized as the ratio of the across-network SD of fitted relative levels after 5 years to that at baseline, with uncertainty from 4,999 multiplier draws (Section~S3).

FP was characterized by its contrast with the equal-weight mean of the other nine networks,
\begin{equation}
    Y^{\mathrm{FP\text{-}rel}}=L^{\mathrm{STD}}_{\mathrm{FP}}-\frac{1}{9}\sum_{k\neq\mathrm{FP}}L^{\mathrm{STD}}_{k}=\frac{10}{9}R^{\mathrm{STD}}_{\mathrm{FP}},
\end{equation}
fitted as a scalar outcome with the four-state design. A 4-df Wald test evaluated the four state slopes jointly, and a 3-df test evaluated heterogeneity across states. To place FP within the broader profile, the four state-specific FP-minus-network differences for each of the nine comparators formed a trajectory contrast. Familywise control across comparators used the maximum Wald statistic over 4,999 shared pedigree-cluster multiplier draws, and equal-state differences used a maximum-$|t|$ procedure. FP's own uncentered signal-normalized level was summarized as an equal-state slope from the joint model, with Bonferroni intervals across the 30 equal-state network-level slopes estimated for the signal-normalized, candidate total-normalized, and attenuation components. FP was characterized after the profile results were obtained in the same cohort.

\subsection{Normalization comparison}

The common-level and joint profile models were refitted to the candidate total-normalized outcomes and to the attenuation term $H$ on the same 852 participants and 8,727 session--state observations, with identical parcel support and the GLS weights of the corresponding signal-normalized models. With shared weights, the difference between representations equals the $H$ coefficients exactly. Paired tests used joint pedigree-level influence functions for the two representations. Holm adjustment was applied within two families:
\begin{itemize}
    \item the three common-level four-state tests (signal-normalized, $H$, and candidate total-normalized);
    \item the six profile omnibus tests (time$\times$network and time$\times$network$\times$state for each of the three components).
\end{itemize}
FP-relative contrasts in this comparison were derived from the joint profile model and therefore differ slightly from the scalar FP model. FP-specific tests were reported without further adjustment.

\subsection{Site, robustness, structure, and cognition}

Site dependence of the profile was evaluated with site$\times$time$\times$network and site$\times$time$\times$state$\times$network tests. From the pooled joint model, we extracted model-implied equal-state profiles for each site and correlated each with the mean profile of the other three, with simultaneous 95\% intervals on the Fisher $z$ scale. For the FP-relative contrast, we tested site$\times$time and site$\times$time$\times$state terms, estimated site-specific equal-state slopes, refitted the model omitting each site in turn, recomputed the contrast after omitting each reference network, and compared released partial covariance with native full covariance fitted to the same rows (848 participants, 2,210 sessions, 6,143 session--state rows), using FP-relative contrasts derived from the ten-network profile models for both.

Network cortical thickness was the surface-area-weighted mean of parcel thickness (at least 95\% parcel coverage), and FP-relative thickness was the FP value minus the mean of the other nine networks. Network volume and surface area were summed across parcels and log transformed, with volume expressed relative to estimated intracranial volume. FP-relative values were defined analogously. Structural trajectories were fitted with baseline age, elapsed time, sex, study, site--scanner, and site$\times$time terms and a participant random-intercept working covariance. To assess whether structure accounted for the fMRI trajectory, each participant's FP-relative thickness was split into between-person (participant mean) and within-person (deviation) components. These were added, with their state interactions, to the FP-relative fMRI model. Adjusted and unadjusted four-state slope vectors were compared on identical rows with a Wald test of their difference, and the change was summarized as the percentage change in the Euclidean norm of the four slopes. The thickness-adjusted model was also refitted in the 822 participants scanned on a single scanner.

For cognition, we used the Fluid Cognition Composite (FluidIQ). Visits were included when all four FP-relative fMRI values and FluidIQ were available at the same event, and participants were required to have at least two such visits (746 participants, 1,881 visits). For each state, the FP-relative value was split into a participant mean and a within-person deviation. FluidIQ was regressed simultaneously on the four between-person and four within-person terms, adjusting for baseline age, elapsed time, sex, study, site--scanner, and site$\times$time, with a participant random-intercept working covariance and pedigree-robust inference. The primary test was the 4-df joint test of the within-person coefficients; between-person coefficients were tested jointly as context. Within-person coefficients were also expressed as standardized regression coefficients using the observed within-person SDs.

Moderation by baseline age was assessed by adding baseline-age$\times$elapsed-time and baseline-age$\times$elapsed-time$\times$state terms (4 df for the FP-relative contrast; 9 df for the equal-state profile).
\section{Results}

\subsection{Overall covariance and signal amplitude declined over follow-up}

The primary cohort included 852 adults with a mean baseline age of 60.96 years (SD 15.12; range 36--89), of whom 492 (57.7\%) were women (Table~\ref{tab:cohort}). Participants contributed 2,220 imaging sessions and 8,727 analyzed session--state rows. Median follow-up was 5.50 years (IQR 3.10--6.33), and 431 participants had at least three sessions.

\begin{table}[!htbp]
\centering
\caption{Cohort characteristics and longitudinal coverage. State coverage counts eligible parent session--state observations before applying the complete ten-network profile criteria; 8,727 session--state observations entered the primary analysis.}
\label{tab:cohort}
\small
\begin{tabularx}{\textwidth}{@{}p{0.22\textwidth}Xr@{}}
\toprule
Category & Measure & Value \\
\midrule
\multicolumn{3}{@{}l}{\textit{Primary imaging cohort}} \\
& Participants & 852 \\
& Sessions & 2,220 \\
& Session--state rows & 8,727 \\
& Participants with $\geq 3$ sessions & 431 \\
& Baseline age, mean (SD), years & 60.96 (15.12) \\
& Baseline age, range, years & 36--89 \\
& Female / male & 492 / 360 \\
\addlinespace
\multicolumn{3}{@{}l}{\textit{Longitudinal coverage}} \\
& Follow-up, median [IQR], years & 5.50 [3.10, 6.33] \\
& Same-scanner / scanner-changed participants & 822 / 30 \\
\addlinespace
\multicolumn{3}{@{}l}{\textit{State coverage}} \\
& REST / CARIT / FaceName / visual--motor sessions & 2,220 / 2,179 / 2,166 / 2,184 \\
\addlinespace
\multicolumn{3}{@{}l}{\textit{Analysis subsets}} \\
& Repeated structural imaging participants & 852 \\
& $\geq 2$ paired FluidIQ visits & 746 \\
& One / no paired FluidIQ visit & 87 / 19 \\
\bottomrule
\end{tabularx}
\end{table}

\begin{figure}[!htbp]
    \centering
    \includegraphics[width=\textwidth]{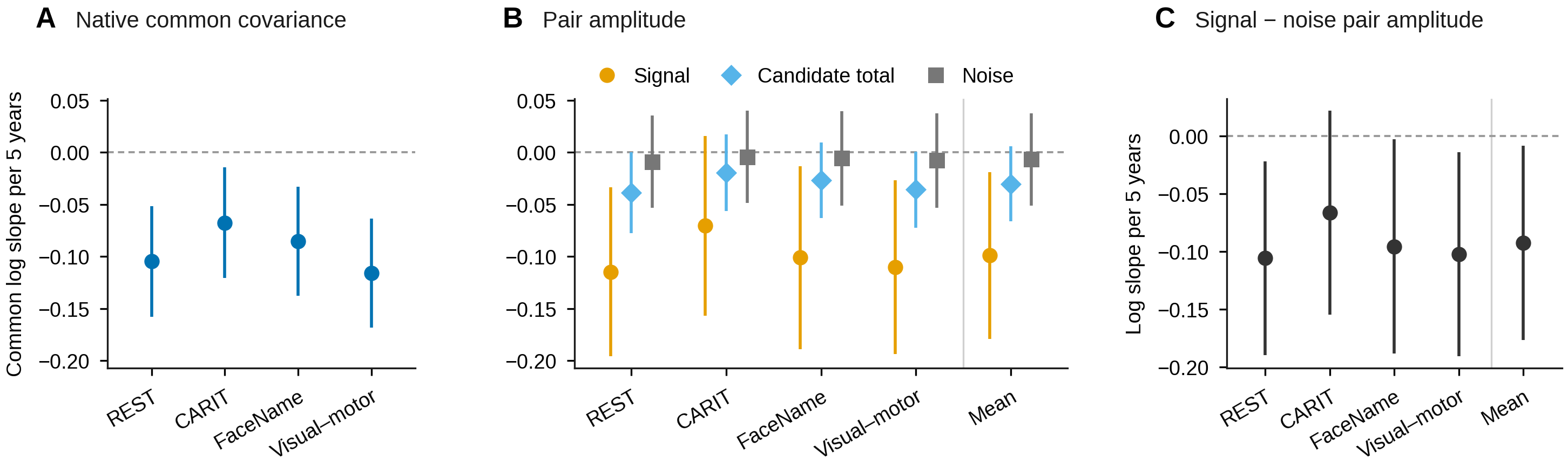}
    \caption{Overall longitudinal change in covariance and amplitude. (A) Native common within-network covariance slopes in each fMRI state. Intervals are simultaneous 95\% intervals across the four states. (B) Slopes of released signal pair amplitude, candidate total pair amplitude, and estimated noise pair amplitude, by state and averaged across states (Mean). The three quantities were fitted with identical weights. Intervals are Bonferroni simultaneous across a family of 20 slopes, which also includes five attenuation-term summaries not shown. Panels A and B share the vertical scale. (C) Paired difference between signal and noise pair-amplitude slopes. Intervals are Bonferroni simultaneous across the five contrasts. Slopes are in log units per 5 years.}
    \label{fig:overall}
\end{figure}

Native common within-network covariance declined in all four states (Fig.~\ref{fig:overall}A). Slopes were $-0.105$ log units per 5 years in REST, $-0.068$ in CARIT, $-0.085$ in FaceName, and $-0.116$ in visual--motor, corresponding to declines of approximately 7--11\% per 5 years ($\chi^2_4=47.69$, $p=1.10\times10^{-9}$).

Released signal pair amplitude declined at a similar rate ($-0.115$, $-0.071$, $-0.101$, and $-0.110$ log units per 5 years; $\chi^2_4=26.09$, Holm-adjusted $p=9.1\times10^{-5}$; Fig.~\ref{fig:overall}B), an equal-state decline of 9.4\% per 5 years (simultaneous 95\% interval 1.9--16.4\%). The estimated noise pair-amplitude slope was small (equal-state $-0.7\%$ per 5 years, simultaneous 95\% interval $-5.0\%$ to $+3.8\%$; $\chi^2_4=2.34$, $p=.673$). In a paired comparison, signal amplitude declined more than noise amplitude (equal-state difference $-0.092$ log units per 5 years, simultaneous 95\% interval $-0.177$ to $-0.008$; Fig.~\ref{fig:overall}C). State-specific estimates are reported in Table~S3B.

\subsection{Network-relative covariance reconfigured, with a recurrent frontoparietal-relative shift}

\begin{figure}[!htbp]
    \centering
    \includegraphics[width=\textwidth]{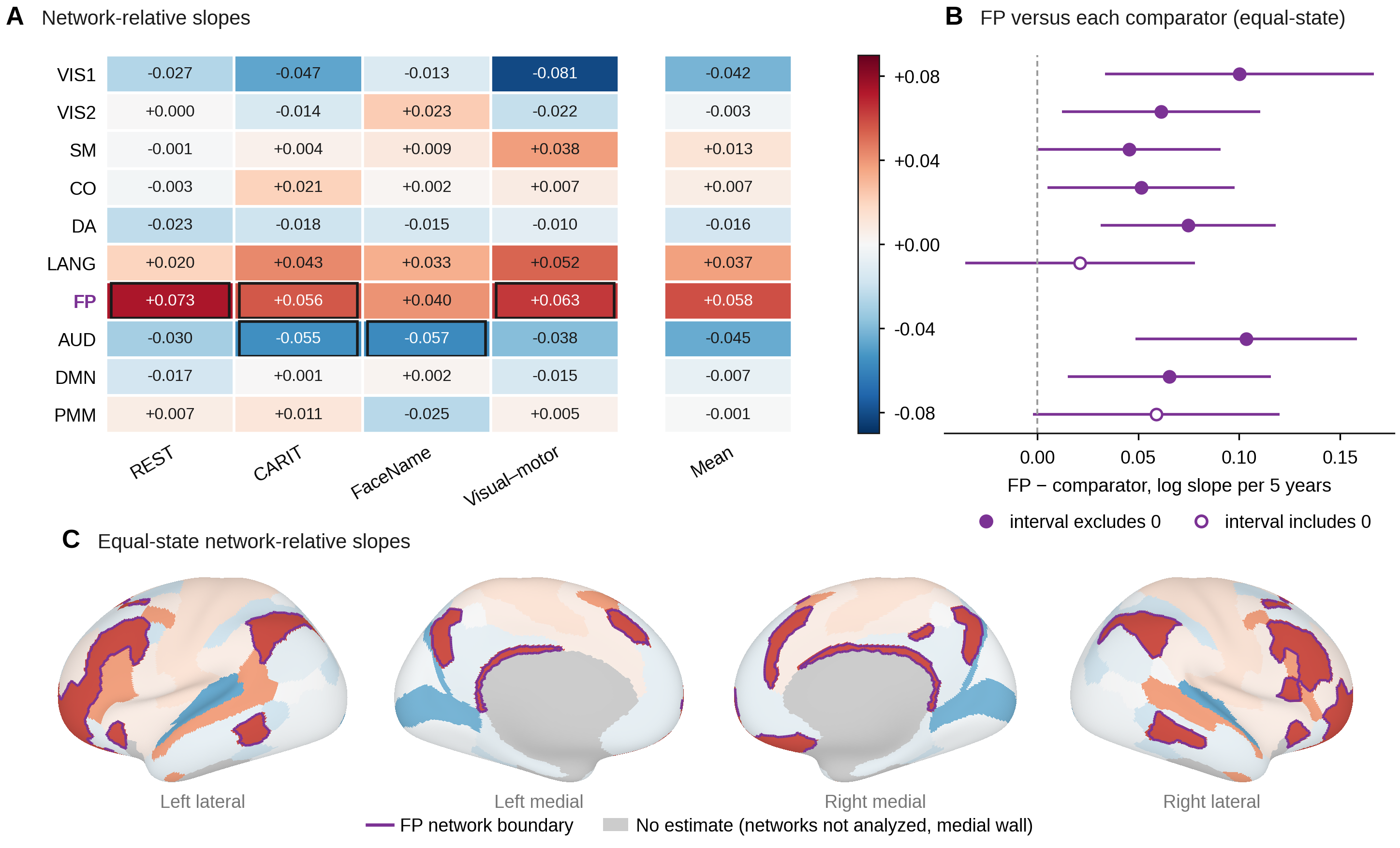}
    \caption{Longitudinal reconfiguration of the network-relative profile. (A) Signal-normalized relative-profile slopes (log units per 5 years) for ten networks in each fMRI state, with the equal-state mean in the right column. Outlined cells have Bonferroni simultaneous intervals across the 40 network--state slopes that exclude zero. Inference on the profile rests on the omnibus tests reported in the text. (B) Equal-state difference between the FP slope and each comparator network, aligned with the rows in A, with max-$|t|$ simultaneous 95\% intervals across the nine comparators. Filled markers indicate intervals excluding zero. (C) Equal-state mean slopes from A shown in anatomical position on the inflated cortical surface, with the same color scale as A. Each network's estimate is assigned uniformly to its HCP-MMP1.0 parcels, so the map displays network-level values, not parcel-level estimates. The purple line marks the FP network boundary. Gray indicates the medial wall and the two networks not analyzed. VIS, visual; SM, somatomotor; CO, cingulo-opercular; DA, dorsal attention; LANG, language; FP, frontoparietal; AUD, auditory; DMN, default mode; PMM, posterior multimodal.}
    \label{fig:network-reconfiguration}
\end{figure}

Network-level analyses used the signal-normalized (amplitude-standardized) relative profile $R^{\mathrm{STD}}$. Section~\ref{sec:normalization} examines which conclusions persist under an alternative normalization. The relative balance among the ten networks changed over follow-up ($\chi^2_9=43.90$, $p=1.47\times10^{-6}$), and the pattern of change differed across states ($\chi^2_{27}=112.00$, $p=2.55\times10^{-12}$; Fig.~\ref{fig:network-reconfiguration}A). Averaged across states, FP had the most positive relative slope ($+0.058$ log units per 5 years), followed by Language ($+0.037$); Auditory ($-0.045$) and Visual1 ($-0.042$) had the most negative. Five of the 40 network--state slopes had Bonferroni simultaneous intervals excluding zero: FP in REST, CARIT, and visual--motor, and Auditory in CARIT and FaceName. The fitted profile also contracted modestly over 5 years (equal-state ratio of across-network SD 0.926, 95\% multiplier interval 0.896--0.960; Table~S4B and Fig.~S2).

The direct contrast between FP and the mean of the other nine networks was positive in every state: $+0.081$ log units per 5 years in REST, $+0.063$ in CARIT, $+0.045$ in FaceName, and $+0.070$ in visual--motor (simultaneous 95\% intervals 0.044--0.118, 0.023--0.102, 0.004--0.085, and 0.025--0.116), or approximately 5--8\% per 5 years ($\chi^2_4=37.08$, $p=1.73\times10^{-7}$). Its magnitude differed across states ($\chi^2_3=12.76$, $p=.0052$). FP's four-state trajectory differed from those of 8 of the 9 other networks after familywise adjustment, and its equal-state slope exceeded those of 7 of 9; the differences from Language and Posterior-Multimodal were not resolved (Fig.~\ref{fig:network-reconfiguration}B; Table~S5). FP's own signal-normalized network level, before centering, also increased (equal-state $+0.070$ log units per 5 years, simultaneous 95\% interval $+0.005$ to $+0.134$; Table~S6), so the relative shift did not arise solely from declining comparator networks.

\subsection{Normalization changed the common-level trajectory but not the frontoparietal direction}
\label{sec:normalization}

\begin{figure}[!htbp]
    \centering
    \includegraphics[width=\textwidth]{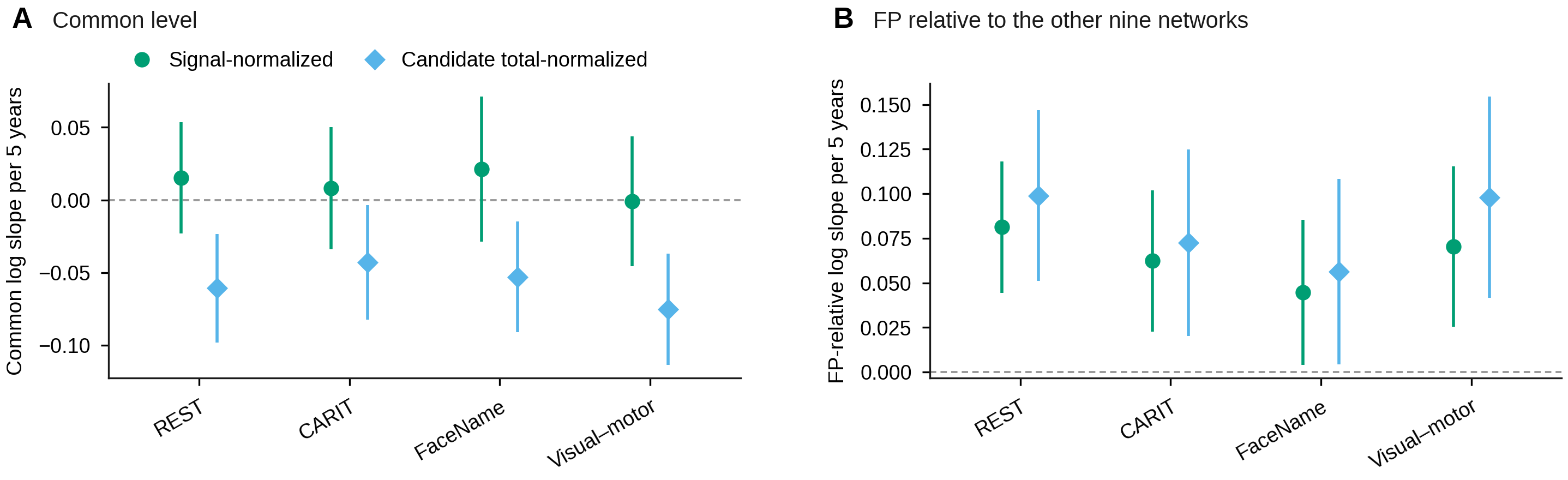}
    \caption{Common-level and FP-relative trajectories under two normalizations. (A) Common-level slopes under signal normalization (released noise-removed amplitude) and candidate total normalization (combined signal and noise variance). (B) FP-relative slopes (FP minus the mean of the other nine networks) under the same two normalizations, estimated from the joint profile model. Signal-normalized values therefore differ slightly from the scalar FP model in Section~3.2. Intervals are Bonferroni simultaneous across the four states within each representation. Slopes are in log units per 5 years. Total amplitude is a candidate reconstruction from released amplitude products (Section~\ref{sec:representations}).}
    \label{fig:normalization}
\end{figure}

We repeated the common-level and network analyses after normalizing covariance by candidate total pair amplitude, which combines the released signal and noise variances (Section~\ref{sec:representations}). Under signal normalization, common-level change was small and not jointly detected (slopes $+0.015$, $+0.008$, $+0.021$, and $-0.001$ log units per 5 years; $\chi^2_4=4.59$, Holm-adjusted $p=.332$). Under candidate total normalization, the common level declined in every state ($-0.061$, $-0.043$, $-0.053$, and $-0.075$; $\chi^2_4=39.26$, Holm-adjusted $p=1.8\times10^{-7}$; Fig.~\ref{fig:normalization}A; Table~S3C). The two trajectories differed ($\chi^2_4=13.98$, Holm-adjusted $p=.015$). By construction, this difference equals the change in the log ratio of signal to candidate total pair amplitude: signal pair amplitude declined more steeply than candidate total pair amplitude (equal-state 9.4\% versus 3.0\% per 5 years), consistent with the signal--noise contrast in Section~3.1. Under signal normalization, the largest state-specific fraction of the native common slope remaining was 0.21, but its simultaneous 95\% upper bound was 1.13, so the prespecified criterion of less than one half remaining in every state was not met (Table~S3A and Fig.~S3).

The positive FP-relative direction, in contrast, persisted. Under candidate total normalization, FP-relative slopes were $+0.099$, $+0.072$, $+0.056$, and $+0.098$ log units per 5 years ($\chi^2_4=34.43$, $p=6.09\times10^{-7}$; Fig.~\ref{fig:normalization}B), exceeding the signal-normalized slopes by 0.010--0.028 (paired $\chi^2_4=5.79$, $p=.215$). Both profile omnibus tests also remained significant (time$\times$network $\chi^2_9=35.98$, Holm-adjusted $p=8.0\times10^{-5}$; time$\times$network$\times$state $\chi^2_{27}=155.62$, Holm-adjusted $p=2.9\times10^{-19}$). Individual network--state slopes did change. For example, Visual1 in CARIT moved from $-0.047$ to $+0.027$ (Fig.~S4 and Table~S7).

\subsection{Site, scanner, and secondary analyses}

The detailed network profile varied across sites, but the equal-state ordering was similar: each site's profile correlated with the mean of the other three ($r=.790$--$.922$; Tables~S8 and Fig.~S5). The equal-state FP-relative slope was positive at every site ($+0.056$ to $+0.077$ log units per 5 years; all simultaneous intervals excluding zero), remained positive in every state when each site was omitted in turn (Fig.~S6), and was positive in all four states in a thickness-adjusted model restricted to the 822 participants scanned on a single scanner (joint $p=2.78\times10^{-7}$; site and scanner composition in Table~S9). Omitting any one network from the nine-network reference changed the equal-state FP-relative slope by at most 0.0054. Under partial covariance, FP-relative slopes remained positive in every state but were smaller on average than those from native full covariance on the same rows (equal-state $+0.066$ versus $+0.127$). The equal-state profile test was not detected ($p=.137$), whereas the state-dependent profile test was ($p=.0029$; Table~S10 and Fig.~S7).

FP cortical thickness declined less than the nine-network reference (FP-relative slope $+0.0156$ mm per 5 years, $p=2.93\times10^{-9}$). Adjusting the FP-relative fMRI model for within- and between-person thickness increased the norm of its four-state slope vector by 2.1\%, and the adjusted and unadjusted slopes did not differ detectably (joint $p=.719$; Table~S11 and Fig.~S8). Moderation of the FP-relative trajectory by baseline age was not detected (joint $p=.748$; Table~S12).

In the 746 participants with at least two visits pairing four-state fMRI with FluidIQ (1,881 visits), within-person coupling between the FP-relative measure and FluidIQ was not detected ($\chi^2_4=5.63$, $p=.228$; standardized coefficients $-0.057$ to $+0.040$, all simultaneous intervals including zero), whereas between-person associations were ($\chi^2_4=36.58$, $p=2.20\times10^{-7}$). This subset was younger than participants outside it (mean baseline age 60.1 versus 67.0 years) and had longer imaging follow-up (Table~S13 and Fig.~S9).
\section{Discussion}

Within individuals followed for a median of 5.5 years, the relative balance of covariance among large-scale cortical networks changed over adult aging, and the frontoparietal network rose relative to the others in rest and in all three task states. This reconfiguration occurred alongside a decline in native covariance of roughly 7--11\% per 5 years, which paralleled a decline in released signal amplitude. Estimated noise amplitude changed much less. Whether the common level of covariance declined depended on how amplitude was normalized, but the positive frontoparietal-relative direction did not. Prior cross-sectional work established that age differences in covariance-based connectivity are closely tied to BOLD amplitude \citep{Strain2022}. The present study extends that work longitudinally, across four brain states, and to the organization of networks relative to one another.

The four-state design separates what recurred from what depended on state. The FP-relative slope was positive in every state, but its magnitude and the full ten-network profile differed across states, so recurrence applies to the direction of the frontoparietal shift rather than to an invariant network pattern. The equal-state point estimates were positive for FP and Language and negative for Auditory and Visual1, and FP's own signal-normalized level increased before centering, so its relative rise was not produced only by declines elsewhere. Previous work has shown greater frontal recruitment in older adults during task performance \citep{Davis2008} and flexible engagement of frontoparietal regions across many tasks \citep{Cole2013}. The present results add a within-person trajectory in frontoparietal covariance, measured relative to other networks, that recurs across both rest and task. FP cortical thickness also declined less than that of the other networks, and the functional trajectory persisted after adjustment for cortical thickness.

These results should be read alongside reports of declining segregation in associative systems, including frontoparietal control networks \citep{Chan2014,Malagurski2020}. The measures differ: segregation contrasts within- and between-network connectivity, whereas the present profile compares within-network covariance across networks, so a relative increase in FP within-network covariance can coexist with declining FP segregation. The fitted profile also contracted modestly in three states, as networks starting low tended to rise and those starting high tended to fall. This partial convergence of network-relative levels is conceptually related to, but distinct from, neural dedifferentiation \citep{KoenRugg2019}.

The normalization results bear on how longitudinal connectivity findings are interpreted. Native covariance declined alongside released signal pair amplitude. The estimated common trajectory was near zero under signal normalization but negative under candidate total normalization, and the two trajectories differed directly. That difference corresponded to a decline in signal pair amplitude relative to candidate total pair amplitude. A correlation-type measure, which retains noise variance in its denominator, can therefore show a common decline that a signal-normalized measure does not. The present longitudinal decline under total normalization differs from the comparative preservation of correlation reported cross-sectionally \citep{Strain2022}. The studies also differ in design, network aggregation, and denominator construction. BOLD amplitude reflects neurovascular as well as neuronal factors, both of which change with age \citep{Liu2013,Tsvetanov2021}. The present decomposition identifies which conclusions depend on amplitude normalization, not the physiological source of the amplitude decline. Network-level conclusions were more stable: both profile omnibus tests remained significant and FP-relative slopes remained positive in every state under both normalizations, while individual network--state estimates changed.

Increased frontal recruitment or connectivity in older adults is often interpreted as compensation or scaffolding \citep{ParkReuterLorenz2009,Cabeza2018}, an interpretation that requires a link to maintained or improved performance. Within-person coupling between the FP-relative measure and fluid cognition was not detected, although between-person associations were. The present data therefore establish the trajectory but not its function. Testing compensation directly will require repeated, task-relevant cognitive measurements that allow within-person associations to be estimated precisely.

Several limitations define the scope of these conclusions. The analysis used released imaging-derived measures, so residual motion and session-level data quality could not be assessed directly, and total amplitude was reconstructed from released products. The signal-normalized measure is a network-level ratio, so changes in relative parcel amplitudes within a network can contribute to it. FP was characterized after the profile results in the same cohort, and its prominence should be confirmed in an independent sample. The detailed profile varied across sites, although the equal-state ordering and the positive FP-relative slope were consistent at every site. Finally, moderation by baseline age was not detected, so the reported effects are averages across the observed age range.

Adult aging, on this evidence, is not well described as uniform weakening of functional networks. Native covariance declines alongside signal pair amplitude, and whether a common decline remains depends on how amplitude is normalized. Beneath that overall change, the relative organization of networks shifts in a state-dependent way, with a positive frontoparietal-relative shift that recurs across rest and task.

\section*{CRediT authorship contribution statement}

\noindent
Conceptualization: T. Liu, W. Dong, H. Zheng, D. Liang, Z. Hu, G. Cheng, and N. Zhang;
Methodology: T. Liu, W. Dong;
Software: T. Liu;
Validation: T. Liu;
Formal analysis: T. Liu;
Investigation: T. Liu;
Visualization: T. Liu;
Writing -- original draft: T. Liu;
Writing -- review \& editing: T. Liu, W. Dong, H. Zheng, D. Liang, Z. Hu, G. Cheng, and N. Zhang;
Supervision: H. Zheng, D. Liang, Z. Hu, G. Cheng, and N. Zhang.

\section*{Funding}
This work was supported by the Shenzhen Science and Technology Program (JCYJ20241202124922031, JCYJ20230807140702006, and KCXFZ20211020163408012) and the Guangdong Basic and Applied Basic Research Foundation of China (2023B1515120007, 2023B1515020002, 2024B1515040018, and 2025B1515120015).

\section*{Declaration of competing interest}
The authors declare that they have no known competing financial interests or personal relationships that could have appeared to influence the work reported in this paper.

\section*{Data and code availability}
The imaging and phenotypic data used in this study were obtained from AABC Release 2 through BALSA (\url{https://balsa.wustl.edu/}). Access is available to eligible researchers under the AABC Data Use Terms. The participant-level data cannot be redistributed by the authors. Analysis code for this study is available at \url{https://github.com/tingshanL/four-state-fmri-aging}.

\section*{Acknowledgements}
Data, methods used, and/or research reported in this publication were provided in whole or in part by the Aging Adult Vulnerability and Resiliency in the Aging Adult Brain Connectome (AABC) project (U19AG073585) and the Human Connectome Project in Aging (HCP-A, U01AG052564) funded by the National Institute on Aging of the National Institutes of Health. HCP-A was further supported by funds provided by the McDonnell Center for Neuroscience at Washington University in St. Louis.

\bibliographystyle{elsarticle-harv}
\bibliography{bib}

@article{Chan2014,
  author  = {Chan, Micaela Y. and Park, Denise C. and Savalia, Neil K. and Petersen, Steven E. and Wig, Gagan S.},
  title   = {Decreased segregation of brain systems across the healthy adult lifespan},
  journal = {Proceedings of the National Academy of Sciences of the United States of America},
  year    = {2014},
  volume  = {111},
  number  = {46},
  pages   = {E4997--E5006},
  doi     = {10.1073/pnas.1415122111}
}

@article{Deery2023,
  author  = {Deery, Hamish A. and Di Paolo, Robert and Moran, Chris and Egan, Gary F. and Jamadar, Sharna D.},
  title   = {The older adult brain is less modular, more integrated, and less efficient at rest: A systematic review of large-scale resting-state functional brain networks in aging},
  journal = {Psychophysiology},
  year    = {2023},
  volume  = {60},
  number  = {1},
  pages   = {e14159},
  doi     = {10.1111/psyp.14159}
}

@article{Malagurski2020,
  author  = {Malagurski, Brigitta and Liem, Franziskus and Oschwald, Jessica and M{\'e}rillat, Susan and J{\"a}ncke, Lutz},
  title   = {Functional dedifferentiation of associative resting state networks in older adults---A longitudinal study},
  journal = {NeuroImage},
  year    = {2020},
  volume  = {214},
  pages   = {116680},
  doi     = {10.1016/j.neuroimage.2020.116680}
}

@article{Hughes2020,
  author  = {Hughes, Colleen and Faskowitz, Joshua and Cassidy, Brittany S. and Sporns, Olaf and Krendl, Anne C.},
  title   = {Aging relates to a disproportionately weaker functional architecture of brain networks during rest and task states},
  journal = {NeuroImage},
  year    = {2020},
  volume  = {209},
  pages   = {116521},
  doi     = {10.1016/j.neuroimage.2020.116521}
}

@article{Varangis2021,
  author  = {Varangis, Eleanna and Habeck, Christian G. and Stern, Yaakov},
  title   = {Task-based functional connectivity in aging: How task and connectivity methodology affect discovery of age effects},
  journal = {Brain and Behavior},
  year    = {2021},
  volume  = {11},
  number  = {1},
  pages   = {e01954},
  doi     = {10.1002/brb3.1954}
}

@article{Strain2022,
  author  = {Strain, Jeremy F. and Brier, Matthew R. and Tanenbaum, Aaron and Gordon, Brian A. and McCarthy, John E. and Dincer, Aylin and Marcus, Daniel S. and Chhatwal, Jasmeer P. and Graff-Radford, Neill R. and Day, Gregory S. and la Foug{\`e}re, Christian and Perrin, Richard J. and Salloway, Stephen and Schofield, Peter R. and Yakushev, Igor and Ikeuchi, Takeshi and V{\"o}glein, Jonathan and Morris, John C. and Benzinger, Tammie L. S. and Bateman, Randall J. and Ances, Beau M.},
  title   = {Covariance-based vs. correlation-based functional connectivity dissociates healthy aging from {Alzheimer} disease},
  journal = {NeuroImage},
  year    = {2022},
  volume  = {261},
  pages   = {119511},
  doi     = {10.1016/j.neuroimage.2022.119511}
}

@article{KoenRugg2019,
  author  = {Koen, Joshua D. and Rugg, Michael D.},
  title   = {Neural Dedifferentiation in the Aging Brain},
  journal = {Trends in Cognitive Sciences},
  year    = {2019},
  volume  = {23},
  number  = {7},
  pages   = {547--559},
  doi     = {10.1016/j.tics.2019.04.012}
}

@article{ParkReuterLorenz2009,
  author  = {Park, Denise C. and Reuter-Lorenz, Patricia},
  title   = {The Adaptive Brain: Aging and Neurocognitive Scaffolding},
  journal = {Annual Review of Psychology},
  year    = {2009},
  volume  = {60},
  pages   = {173--196},
  doi     = {10.1146/annurev.psych.59.103006.093656}
}

@article{Cabeza2018,
  author  = {Cabeza, Roberto and Albert, Marilyn and Belleville, Sylvie and Craik, Fergus I. M. and Duarte, Audrey and Grady, Cheryl L. and Lindenberger, Ulman and Nyberg, Lars and Park, Denise C. and Reuter-Lorenz, Patricia A. and Rugg, Michael D. and Steffener, Jason and Rajah, M. Natasha},
  title   = {Maintenance, reserve and compensation: the cognitive neuroscience of healthy ageing},
  journal = {Nature Reviews Neuroscience},
  year    = {2018},
  volume  = {19},
  number  = {11},
  pages   = {701--710},
  doi     = {10.1038/s41583-018-0068-2}
}

@article{Liu2013,
  author  = {Liu, Thomas T.},
  title   = {Neurovascular factors in resting-state functional {MRI}},
  journal = {NeuroImage},
  year    = {2013},
  volume  = {80},
  pages   = {339--348},
  doi     = {10.1016/j.neuroimage.2013.04.071}
}

@article{Tsvetanov2021,
  author  = {Tsvetanov, Kamen A. and Henson, Richard N. A. and Rowe, James B.},
  title   = {Separating vascular and neuronal effects of age on {fMRI} {BOLD} signals},
  journal = {Philosophical Transactions of the Royal Society B: Biological Sciences},
  year    = {2021},
  volume  = {376},
  number  = {1815},
  pages   = {20190631},
  doi     = {10.1098/rstb.2019.0631}
}

@article{Glasser2016,
  author  = {Glasser, Matthew F. and Coalson, Timothy S. and Robinson, Emma C. and Hacker, Carl D. and Harwell, John and Yacoub, Essa and Ugurbil, Kamil and Andersson, Jesper and Beckmann, Christian F. and Jenkinson, Mark and Smith, Stephen M. and Van Essen, David C.},
  title   = {A multi-modal parcellation of human cerebral cortex},
  journal = {Nature},
  year    = {2016},
  volume  = {536},
  pages   = {171--178},
  doi     = {10.1038/nature18933}
}

@article{Ji2019,
  author  = {Ji, Jie Lisa and Spronk, Marjolein and Kulkarni, Kaustubh and Repov{\v{s}}, Grega and Anticevic, Alan and Cole, Michael W.},
  title   = {Mapping the human brain's cortical-subcortical functional network organization},
  journal = {NeuroImage},
  year    = {2019},
  volume  = {185},
  pages   = {35--57},
  doi     = {10.1016/j.neuroimage.2018.10.006}
}

@article{Bookheimer2019,
  author  = {Bookheimer, Susan Y. and Salat, David H. and Terpstra, Melissa and Ances, Beau M. and Barch, Deanna M. and Buckner, Randy L. and Burgess, Gregory C. and Curtiss, Sandra W. and Diaz-Santos, Mirella and Elam, Jennifer Stine and Fischl, Bruce and Greve, Douglas N. and Hagy, Hannah A. and Harms, Michael P. and Hatch, Olivia M. and Hedden, Trey and Hodge, Cynthia and Japardi, Kevin C. and Kuhn, Taylor P. and Ly, Timothy K. and Smith, Stephen M. and Somerville, Leah H. and Ugurbil, Kamil and van der Kouwe, Andre and Van Essen, David C. and Woods, Roger P. and Yacoub, Essa},
  title   = {The {Lifespan Human Connectome Project in Aging}: An overview},
  journal = {NeuroImage},
  year    = {2019},
  volume  = {185},
  pages   = {335--348},
  doi     = {10.1016/j.neuroimage.2018.10.009}
}

@article{Harms2018,
  author  = {Harms, Michael P. and Somerville, Leah H. and Ances, Beau M. and Andersson, Jesper and Barch, Deanna M. and Bastiani, Matteo and Bookheimer, Susan Y. and Brown, Timothy B. and Buckner, Randy L. and Burgess, Gregory C. and Coalson, Timothy S. and Chappell, Michael A. and Dapretto, Mirella and Douaud, Gwena{\"e}lle and Fischl, Bruce and Glasser, Matthew F. and Greve, Douglas N. and Hodge, Cynthia and Jamison, Keith W. and Jbabdi, Saad and Kandala, Sridhar and Li, Xiufeng and Mair, Ross W. and Mangia, Silvia and Marcus, Daniel and Mascali, Daniele and Moeller, Steen and Nichols, Thomas E. and Robinson, Emma C. and Salat, David H. and Smith, Stephen M. and Sotiropoulos, Stamatios N. and Terpstra, Melissa and Thomas, Kathleen M. and Tisdall, M. Dylan and Ugurbil, Kamil and van der Kouwe, Andre and Woods, Roger P. and Z{\"o}llei, Lilla and Van Essen, David C. and Yacoub, Essa},
  title   = {Extending the {Human Connectome Project} across ages: Imaging protocols for the Lifespan Development and Aging projects},
  journal = {NeuroImage},
  year    = {2018},
  volume  = {183},
  pages   = {972--984},
  doi     = {10.1016/j.neuroimage.2018.09.060}
}

@misc{AABCRelease2,
  author       = {{Human Connectome Project}},
  title        = {{Aging Adult Brain Connectome Release 2}},
  year         = {2026},
  howpublished = {HCP Aging / Aging Adult Brain Connectome data release},
  url = {https://www.humanconnectome.org/study/hcp-lifespan-aging/document/aging-adult-brain-connectome-release-2},
  note         = {Release date: January 28, 2026}
}

@article{Glasser2013,
  author  = {Glasser, Matthew F. and Sotiropoulos, Stamatios N. and Wilson, J. Anthony and Coalson, Timothy S. and Fischl, Bruce and Andersson, Jesper L. and Xu, Junqian and Jbabdi, Saad and Webster, Matthew and Polimeni, Jonathan R. and Van Essen, David C. and Jenkinson, Mark},
  title   = {The minimal preprocessing pipelines for the {Human Connectome Project}},
  journal = {NeuroImage},
  year    = {2013},
  volume  = {80},
  pages   = {105--124},
  doi     = {10.1016/j.neuroimage.2013.04.127}
}

@article{SalimiKhorshidi2014,
  author  = {Salimi-Khorshidi, Gholamreza and Douaud, Gwena{\"e}lle and Beckmann, Christian F. and Glasser, Matthew F. and Griffanti, Ludovica and Smith, Stephen M.},
  title   = {Automatic denoising of functional {MRI} data: Combining independent component analysis and hierarchical fusion of classifiers},
  journal = {NeuroImage},
  year    = {2014},
  volume  = {90},
  pages   = {449--468},
  doi     = {10.1016/j.neuroimage.2013.11.046}
}

@article{Glasser2018,
  author  = {Glasser, Matthew F. and Coalson, Timothy S. and Bijsterbosch, Janine D. and Harrison, Samuel J. and Harms, Michael P. and Anticevic, Alan and Van Essen, David C. and Smith, Stephen M.},
  title   = {Using temporal {ICA} to selectively remove global noise while preserving global signal in functional {MRI} data},
  journal = {NeuroImage},
  year    = {2018},
  volume  = {181},
  pages   = {692--717},
  doi     = {10.1016/j.neuroimage.2018.04.076}
}

@article{Aitchison1982,
  author  = {Aitchison, John},
  title   = {The statistical analysis of compositional data},
  journal = {Journal of the Royal Statistical Society: Series B (Methodological)},
  year    = {1982},
  volume  = {44},
  number  = {2},
  pages   = {139--160},
  doi     = {10.1111/j.2517-6161.1982.tb01195.x}
}

@article{Egozcue2003,
  author  = {Egozcue, J. J. and Pawlowsky-Glahn, V. and Mateu-Figueras, G. and Barcel{\'o}-Vidal, C.},
  title   = {Isometric logratio transformations for compositional data analysis},
  journal = {Mathematical Geology},
  year    = {2003},
  volume  = {35},
  number  = {3},
  pages   = {279--300},
  doi     = {10.1023/A:1023818214614}
}

@article{Cole2013,
  author  = {Cole, Michael W. and Reynolds, Jeremy R. and Power, Jonathan D. and Repovs, Grega and Anticevic, Alan and Braver, Todd S.},
  title   = {Multi-task connectivity reveals flexible hubs for adaptive task control},
  journal = {Nature Neuroscience},
  year    = {2013},
  volume  = {16},
  number  = {9},
  pages   = {1348--1355},
  doi     = {10.1038/nn.3470}
}

@article{Davis2008,
  author  = {Davis, Simon W. and Dennis, Nancy A. and Daselaar, Sander M. and Fleck, Mathias S. and Cabeza, Roberto},
  title   = {Qu{\'e} {PASA}? {T}he posterior--anterior shift in aging},
  journal = {Cerebral Cortex},
  year    = {2008},
  volume  = {18},
  number  = {5},
  pages   = {1201--1209},
  doi     = {10.1093/cercor/bhm155}
}
\end{document}


\begin{center}
{\Large\bfseries Supplementary Material}\\[0.5em]
{\normalsize
Longitudinal fMRI reveals multi-network reconfiguration and a recurrent
frontoparietal shift across rest and task states during adult aging}
\end{center}
\vspace{1em}


\noindent Notation follows the main text. $L_k$ is the log covariance level of network $k$; $G=\frac{1}{10}\sum_k L_k$ is the common level and $R_k=L_k-G$ the network-relative coordinate. Signal normalization divides network covariance by the mean released signal pair amplitude $P_k$; candidate total normalization uses $t_p=\sqrt{a_p^2+n_p^2}$ and $P_k^{\mathrm{TOT}}$, so that $L_k^{\mathrm{TOT}}=L_k^{\mathrm{STD}}+H_k$ with $H_k=\log(P_k/P_k^{\mathrm{TOT}})\leq0$. Slopes are change in log units per 5 years of elapsed follow-up. $\beta^{\mathrm{RAW}}$, $\beta^{\mathrm{A}}$ and $\beta^{\mathrm{STD}}$ denote native, pair-amplitude and signal-normalized slopes. Unless stated otherwise, uncertainty comes from pedigree-clustered sandwich covariance matrices, and resampling-based intervals use 4,999 Gaussian pedigree-cluster multiplier draws of the fitted coefficient vector, not refits of the model.

\section{Construction of the released covariance and amplitude products}
\label{sec:supp-measurement}

\begin{figure}[!htbp]
    \centering
    \includegraphics[width=\textwidth]{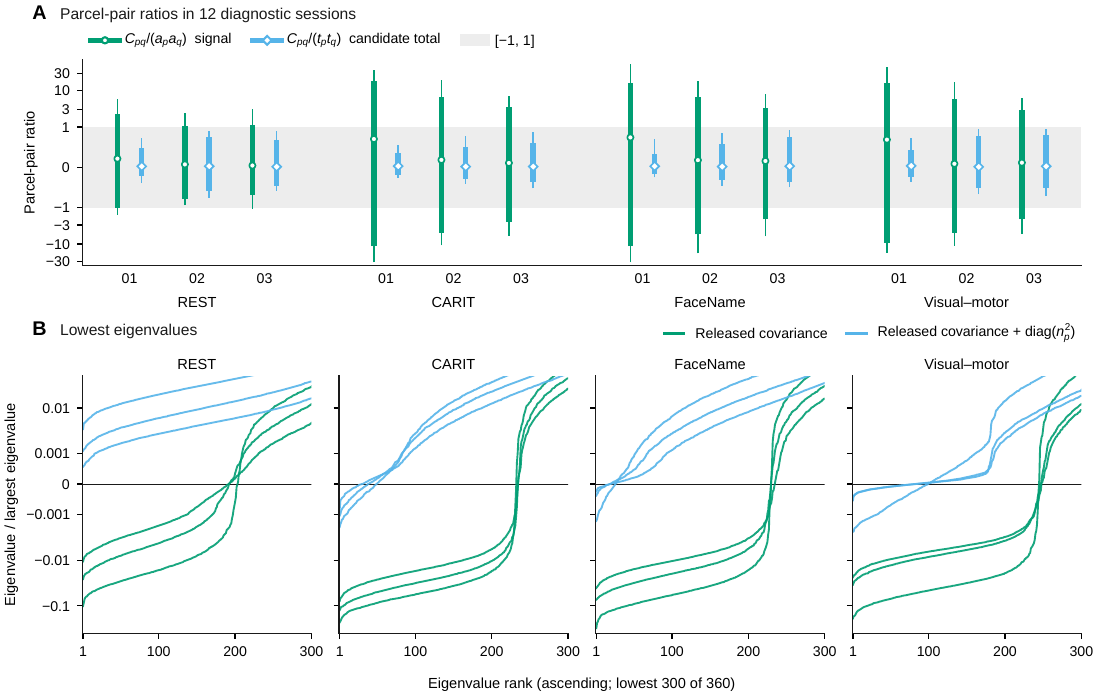}
    \caption{Released covariance construction in 12 diagnostic sessions. (A) Distribution of parcel-pair ratios after dividing released covariance by signal amplitudes, $C_{pq}/(a_pa_q)$ (green), or by candidate total amplitudes, $C_{pq}/(t_pt_q)$ (blue). Thick bars span the 0.1th--99.9th percentiles, thin lines the full range, and open symbols mark medians; the gray band is $[-1,1]$. The vertical axis is linear within $\pm1$ and logarithmic beyond. (B) The lowest 300 of 360 eigenvalues of each released covariance matrix (green) and of the same matrix with the released noise variances added to its diagonal (blue), divided by the largest eigenvalue; three sessions per state. The vertical axis is linear within $\pm10^{-3}$ and logarithmic beyond.}
    \label{fig:supp-measurement}
\end{figure}

Twelve diagnostic sessions (three per state) were used to check how the released full covariance, signal amplitude and noise amplitude products relate to one another (Table~\ref{tab:supp-measurement}; Fig.~\ref{fig:supp-measurement}). Recomputing within-network covariance from the released parcel-by-parcel matrices reproduced the released network summaries (120 network--session values; maximum relative difference $5.2\times10^{-8}$). The diagonal of each released covariance matrix reproduced the squared released signal amplitude $a_p^2$ (median relative error $1.9\times10^{-7}$; maximum $1.4\times10^{-6}$). Dividing off-diagonal covariance by signal amplitudes did not yield bounded correlations: $|C_{pq}/(a_pa_q)|$ exceeded 1.01 in 0.1--3.6\% of pairs in REST sessions and in 7.7--60.5\% of pairs in task sessions. Released matrices had 191--247 negative eigenvalues of 360, with the smallest reaching 1.1--31\% of the largest eigenvalue in magnitude.

These properties are consistent with removal of unstructured noise from the diagonal only. Adding the released noise variance, $\mathrm{diag}(n_p^2)$, made every diagnostic pair ratio $C_{pq}/(t_pt_q)$ fall within $[-0.76,\,0.96]$ and restored positive semidefiniteness in the three REST sessions. Task matrices retained 17--99 small negative eigenvalues, none larger in magnitude than 0.24\% of the largest eigenvalue. The task runs are short relative to the number of parcels (194--345 frames for 360 parcels), so their sample covariance matrices are rank deficient, and the remaining negative eigenvalues were of the order of $\max_p n_p^2/T$ (0.7--2.0 times that value). A small inconsistency between the covariance and noise-variance conventions in rank-deficient matrices would be sufficient to produce them; we report them without repair.

Across the analyzed cohort, the signal-to-candidate-total pair-amplitude ratio $P_k/P_k^{\mathrm{TOT}}$ was smaller in task states than in REST (Table~\ref{tab:supp-attenuation}). Network medians ranged from 0.22 to 0.60 in REST and from 0.09 to 0.53 in the task states, corresponding to differences of roughly 1.7- to 12-fold between signal and candidate total normalization of the same covariance.

\begin{table}[!htbp]
\centering
\caption{Measurement diagnostics in 12 diagnostic sessions (three per state). Each session has 360 parcels and 64,620 parcel pairs. $a_p$, released signal amplitude; $n_p$, released noise amplitude; $t_p=\sqrt{a_p^2+n_p^2}$, candidate total amplitude. $T$ is the nominal number of frames for task runs (Harms et al., 2018); the REST aggregation was not inferred. $n_{<0}$, number of negative eigenvalues (of 360); $\lambda_{\min}/\lambda_{\max}$, smallest eigenvalue divided by the largest. Diagnostic sessions describe the construction of the released products and are not cohort prevalence estimates.}
\label{tab:supp-measurement}
\footnotesize
\begin{adjustbox}{max width=\textwidth}
\begin{tabular}{lrrrrrrrr}
\toprule
Session & $T$ & \multicolumn{2}{c}{$C_{pq}/(a_pa_q)$} & $C_{pq}/(t_pt_q)$ & \multicolumn{2}{c}{Released $C$} & \multicolumn{2}{c}{$C+\mathrm{diag}(n_p^2)$} \\
\cmidrule(lr){3-4}\cmidrule(lr){5-5}\cmidrule(lr){6-7}\cmidrule(lr){8-9}
 & & Range & \% $|\cdot|>1.01$ & Range & $n_{<0}$ & $\lambda_{\min}/\lambda_{\max}$ & $n_{<0}$ & $\lambda_{\min}/\lambda_{\max}$ \\
\midrule
REST 1 & -- & [$-1.58$, $6.08$] & 3.6 & [$-0.39$, $0.74$] & 202 & $-0.1008$ & 0 & $0.0035$ \\
REST 2 & -- & [$-0.94$, $2.36$] & 0.1 & [$-0.76$, $0.90$] & 191 & $-0.0106$ & 0 & $0.0006$ \\
REST 3 & -- & [$-1.09$, $3.06$] & 0.2 & [$-0.58$, $0.91$] & 192 & $-0.0259$ & 0 & $0.0012$ \\
\addlinespace[2pt]
CARIT 1 & 300 & [$-31.39$, $37.67$] & 57.8 & [$-0.26$, $0.55$] & 232 & $-0.2307$ & 48 & $-0.0019$ \\
CARIT 2 & 300 & [$-10.96$, $19.23$] & 26.6 & [$-0.40$, $0.78$] & 234 & $-0.1246$ & 38 & $-0.0011$ \\
CARIT 3 & 300 & [$-5.86$, $6.99$] & 9.3 & [$-0.52$, $0.86$] & 234 & $-0.0787$ & 29 & $-0.0008$ \\
\addlinespace[2pt]
FaceName 1 & 345 & [$-31.65$, $55.38$] & 60.5 & [$-0.25$, $0.70$] & 229 & $-0.3067$ & 26 & $-0.0014$ \\
FaceName 2 & 345 & [$-18.30$, $18.27$] & 22.1 & [$-0.45$, $0.85$] & 229 & $-0.0731$ & 17 & $-0.0004$ \\
FaceName 3 & 345 & [$-6.20$, $8.23$] & 7.7 & [$-0.49$, $0.93$] & 234 & $-0.0407$ & 18 & $-0.0002$ \\
\addlinespace[2pt]
Visual--motor 1 & 194 & [$-17.56$, $44.12$] & 58.6 & [$-0.35$, $0.72$] & 244 & $-0.1876$ & 99 & $-0.0024$ \\
Visual--motor 2 & 194 & [$-11.33$, $17.72$] & 19.5 & [$-0.66$, $0.95$] & 247 & $-0.0345$ & 83 & $-0.0005$ \\
Visual--motor 3 & 194 & [$-5.39$, $6.17$] & 7.9 & [$-0.71$, $0.96$] & 247 & $-0.0236$ & 81 & $-0.0004$ \\
\bottomrule
\end{tabular}
\end{adjustbox}
\end{table}

\begin{table}[!htbp]
\centering
\caption{Cohort distribution of the network signal-to-candidate-total pair-amplitude ratio $P_k/P_k^{\mathrm{TOT}}=\exp(H_k)$. Entries are the median [5th, 95th percentile] across matched session--state observations (852 participants). The ratio compares mean parcel-pair amplitude products; it is not a mean parcel signal-variance fraction.}
\label{tab:supp-attenuation}
\footnotesize
\begin{adjustbox}{max width=\textwidth}
\begin{tabular}{lcccc}
\toprule
Network & REST ($n=2{,}220$) & CARIT ($n=2{,}179$) & FaceName ($n=2{,}165$) & Visual--motor ($n=2{,}163$) \\
\midrule
Visual1 & 0.52 [0.25, 0.81] & 0.21 [0.04, 0.51] & 0.33 [0.08, 0.58] & 0.43 [0.15, 0.68] \\
Visual2 & 0.58 [0.31, 0.82] & 0.27 [0.06, 0.52] & 0.41 [0.10, 0.65] & 0.53 [0.22, 0.73] \\
Somatomotor & 0.46 [0.20, 0.74] & 0.19 [0.05, 0.45] & 0.22 [0.06, 0.48] & 0.27 [0.09, 0.53] \\
Cingulo-Opercular & 0.37 [0.19, 0.54] & 0.15 [0.04, 0.34] & 0.17 [0.05, 0.34] & 0.24 [0.08, 0.44] \\
Dorsal-attention & 0.60 [0.36, 0.77] & 0.29 [0.08, 0.52] & 0.33 [0.08, 0.54] & 0.44 [0.15, 0.66] \\
Language & 0.43 [0.20, 0.64] & 0.17 [0.05, 0.38] & 0.23 [0.07, 0.42] & 0.21 [0.07, 0.43] \\
Frontoparietal & 0.53 [0.31, 0.69] & 0.26 [0.06, 0.49] & 0.33 [0.08, 0.55] & 0.29 [0.09, 0.52] \\
Auditory & 0.22 [0.07, 0.50] & 0.09 [0.02, 0.26] & 0.10 [0.03, 0.28] & 0.11 [0.03, 0.28] \\
Default & 0.38 [0.21, 0.54] & 0.18 [0.04, 0.41] & 0.17 [0.04, 0.33] & 0.21 [0.07, 0.42] \\
Posterior-Multimodal & 0.52 [0.24, 0.79] & 0.20 [0.05, 0.46] & 0.24 [0.07, 0.49] & 0.31 [0.09, 0.57] \\
\bottomrule
\end{tabular}
\end{adjustbox}
\end{table}

\section{Common level and amplitude components}
\label{sec:supp-common}

Native common covariance declined in all four states ($\chi^2_4=47.69$, $p=1.10\times10^{-9}$), and the separately fitted pair-amplitude component declined in parallel ($\chi^2_4=25.18$, $p=4.62\times10^{-5}$), whereas the signal-normalized common level was not jointly different from zero ($\chi^2_4=4.59$, $p=.332$; Table~\ref{tab:supp-common}A and C). Refitting the two components with the native model's weights reproduced $\beta^{\mathrm{RAW}}_s=\beta^{\mathrm{A}}_s+\beta^{\mathrm{STD}}_s$ to machine precision. The residual fraction $q_s$ ranged from 0.045 to 0.206. Its simultaneous upper bounds ranged from 0.41 to 1.13, so the prespecified criterion (upper bound below 0.50 in every state) was not met. This ratio is imprecise because its denominator is estimated; the native slope changed sign in 0.1\% of CARIT draws and in no draws for the other states.

Refitted with the signal-normalized model's weights, the signal pair amplitude declined by 6.8--10.8\% per 5 years across states (equal-state mean $-9.4$\%), the candidate total pair amplitude by 1.9--3.8\%, and the estimated noise pair amplitude by 0.4--0.9\% (Table~\ref{tab:supp-common}B). Signal and total, but not noise, slopes were jointly different from zero after Holm adjustment. In the directly paired contrast, signal pair amplitude declined more than noise pair amplitude (equal-state difference $-0.092$ [$-0.177$, $-0.008$]); the state-specific intervals excluded zero except in CARIT.

Under candidate total normalization, the common level declined in every state ($-0.043$ to $-0.075$; $\chi^2_4=39.26$, Holm $p=1.84\times10^{-7}$; Table~\ref{tab:supp-common}C), corresponding to 4.2--7.2\% per 5 years. The difference between the two normalizations is the attenuation term $H$, which declined by 0.051--0.076 per 5 years (Holm $p=.015$). Because $L^{\mathrm{TOT}}=L^{\mathrm{STD}}+H$ holds observation by observation and the three outcomes share fixed weights, the paired slope differences equal the $H$ slopes exactly. Site$\times$time interactions in the common level were nominally detected under each normalization (Table~\ref{tab:supp-common}A and C) and are reported as descriptive.

\begin{table}[!htbp]
\centering
\caption{Common-level longitudinal slopes (log units per 5 years) with simultaneous 95\% intervals. Panel A: native common slope and the separately fitted pair-amplitude component (Bonferroni across four states); $\beta^{\mathrm{A}}$ and $\beta^{\mathrm{STD}}$ are refits of the two components with the native model's GLS weights, so that $\beta^{\mathrm{RAW}}_s=\beta^{\mathrm{A}}_s+\beta^{\mathrm{STD}}_s$ exactly; $q_s=|\beta^{\mathrm{STD}}_s/\beta^{\mathrm{RAW}}_s|$ with a max-$|t|$ simultaneous band from 4,999 pedigree-cluster multiplier draws. Panel B: log pair-amplitude products refitted with the signal-normalized common model's weights (Bonferroni across 20 slopes); signal $-$ noise is the directly paired contrast (Bonferroni across five summaries); Holm adjustment across the three amplitude omnibus tests. Panel C: signal-normalized and candidate total-normalized common levels, $H=\log(P/P^{\mathrm{TOT}})$, and the identity $L^{\mathrm{TOT}}=L^{\mathrm{STD}}+H$ (Bonferroni across four states within each column); Holm adjustment across the three common-level tests. Site interaction tests (3 and 9 df) are nominal and descriptive.}
\label{tab:supp-common}
\footnotesize
\begin{adjustbox}{max width=\textwidth}
\begin{tabular}{llllll}
\toprule
\multicolumn{6}{l}{\textit{A. Native common slope and its fixed-weight decomposition (native-model weights)}} \\
State & Native $\beta^{\mathrm{RAW}}$ & Pair amplitude, own weights & $\beta^{\mathrm{A}}$ & $\beta^{\mathrm{STD}}$ & $q_s$ \\
\midrule
REST & $-0.105$ [$-0.158$, $-0.051$] & $-0.1161$ [$-0.1847$, $-0.0476$] & $-0.116$ & $+0.011$ & $0.106$ [$0.000$, $0.497$] \\
CARIT & $-0.068$ [$-0.121$, $-0.015$] & $-0.0721$ [$-0.1447$, $+0.0005$] & $-0.072$ & $+0.004$ & $0.066$ [$0.000$, $1.134$] \\
FaceName & $-0.085$ [$-0.138$, $-0.033$] & $-0.1031$ [$-0.1767$, $-0.0294$] & $-0.103$ & $+0.018$ & $0.206$ [$0.000$, $0.950$] \\
Visual--motor & $-0.116$ [$-0.168$, $-0.064$] & $-0.1115$ [$-0.1819$, $-0.0410$] & $-0.111$ & $-0.005$ & $0.045$ [$0.000$, $0.408$] \\
Site$\times$time; site$\times$time$\times$state ($p$) & $.0054$; $1.58\times10^{-5}$ & & & &  \\
\midrule
\multicolumn{6}{l}{\textit{B. Pair amplitudes (signal-model weights)}} \\
State & Signal $\log P$ & Candidate total $\log P^{\mathrm{TOT}}$ & Noise $\log P^{\mathrm{N}}$ & Signal $-$ noise & \\
\midrule
REST & $-0.115$ [$-0.196$, $-0.034$] & $-0.039$ [$-0.077$, $-0.0001$] & $-0.009$ [$-0.053$, $+0.035$] & $-0.106$ [$-0.190$, $-0.022$] &  \\
CARIT & $-0.071$ [$-0.157$, $+0.016$] & $-0.020$ [$-0.056$, $+0.017$] & $-0.004$ [$-0.049$, $+0.040$] & $-0.066$ [$-0.155$, $+0.022$] &  \\
FaceName & $-0.101$ [$-0.189$, $-0.013$] & $-0.027$ [$-0.063$, $+0.009$] & $-0.006$ [$-0.051$, $+0.040$] & $-0.096$ [$-0.188$, $-0.003$] &  \\
Visual--motor & $-0.110$ [$-0.194$, $-0.027$] & $-0.036$ [$-0.073$, $+0.001$] & $-0.008$ [$-0.053$, $+0.037$] & $-0.102$ [$-0.191$, $-0.014$] &  \\
Equal-state mean & $-0.099$ [$-0.179$, $-0.019$] & $-0.030$ [$-0.066$, $+0.006$] & $-0.007$ [$-0.051$, $+0.038$] & $-0.092$ [$-0.177$, $-0.008$] &  \\
Four-state $\chi^2_4$ ($p_{\mathrm{Holm}}$) & $26.09$ ($9.10\times10^{-5}$) & $24.36$ ($1.35\times10^{-4}$) & $2.34$ ($.673$) & &  \\
\midrule
\multicolumn{6}{l}{\textit{C. Common level under signal and candidate total normalization (signal-model weights)}} \\
State & Signal-normalized & $H$ & Candidate total-normalized & & \\
\midrule
REST & $+0.015$ [$-0.023$, $+0.053$] & $-0.076$ [$-0.135$, $-0.017$] & $-0.061$ [$-0.098$, $-0.024$] &  &  \\
CARIT & $+0.008$ [$-0.034$, $+0.050$] & $-0.051$ [$-0.115$, $+0.013$] & $-0.043$ [$-0.082$, $-0.004$] &  &  \\
FaceName & $+0.021$ [$-0.029$, $+0.071$] & $-0.074$ [$-0.141$, $-0.007$] & $-0.053$ [$-0.091$, $-0.015$] &  &  \\
Visual--motor & $-0.001$ [$-0.046$, $+0.044$] & $-0.074$ [$-0.137$, $-0.012$] & $-0.075$ [$-0.114$, $-0.037$] &  &  \\
Four-state $\chi^2_4$ ($p_{\mathrm{Holm}}$) & $4.59$ ($.332$) & $13.98$ ($.0147$) & $39.26$ ($1.84\times10^{-7}$) & &  \\
Site$\times$time; site$\times$time$\times$state ($p$) & $7.15\times10^{-4}$; $.0170$ & $2.45\times10^{-6}$; $.0135$ & $1.60\times10^{-4}$; $5.64\times10^{-4}$ & &  \\
\bottomrule
\end{tabular}
\end{adjustbox}
\end{table}

\section{Network profile and FP characterization}
\label{sec:supp-profile}

The complete signal-normalized ten-network profile is given in Table~\ref{tab:supp-profile}A and displayed in main-text Fig.~3. Five of the 40 network--state intervals excluded zero: FP was positive in REST, CARIT and visual--motor, and Auditory was negative in CARIT and FaceName. Equal-state slopes ranged from $-0.045$ (Auditory) and $-0.042$ (Visual1) to $+0.037$ (Language) and $+0.058$ (FP). Across the ten networks, parcel count was not consistently related to the relative slope (Pearson $r=.25$, $.28$, $.63$ and $.22$ in REST, CARIT, FaceName and visual--motor; all $p\geq.052$).

\begin{figure}[!htbp]
    \centering
    \includegraphics[width=\textwidth]{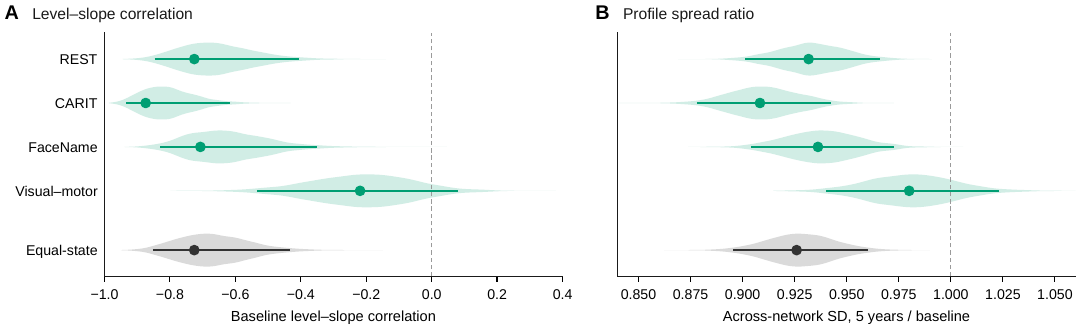}
    \caption{Fitted-profile convergence. (A) Correlation between the fitted baseline relative level and the longitudinal slope across the ten networks. (B) Ratio of the across-network SD of the fitted profile after 5 years to its baseline value. Points are observed values, horizontal lines 95\% intervals, and shaded densities the 4,999 pedigree-cluster multiplier draws. Dashed lines mark 0 and 1.}
    \label{fig:supp-convergence}
\end{figure}

Summaries of the fitted profile supported partial contraction (Table~\ref{tab:supp-profile}B; Fig.~\ref{fig:supp-convergence}). Networks with lower baseline relative levels had more positive slopes in REST, CARIT and FaceName, and the across-network SD after 5 years was 0.908--0.936 times its baseline value in these states. Both summaries were weaker in visual--motor, where the intervals included zero and one, respectively. For the equal-state average, the level--slope correlation was $r=-.725$ [$-.852$, $-.431$] and the SD ratio 0.926 [0.896, 0.960]. FP lay above the fitted level--slope trend (equal-state residual $+0.024$ [$0.002$, $0.046$]), with state-specific intervals excluding zero in REST and visual--motor.

\begin{table}[!htbp]
\centering
\caption{Signal-normalized ten-network profile. Panel A: slopes of the centered log-ratio coordinates $R_k$ (log units per 5 years); $^{*}$ interval excludes zero. Joint tests: time$\times$network $\chi^2_{9}=43.90$, $p=1.47\times10^{-6}$; time$\times$network$\times$state $\chi^2_{27}=112.00$, $p=2.55\times10^{-12}$. Panel B: summaries of the fitted profile recomputed in 4,999 pedigree-cluster multiplier draws of the full coefficient vector; SD is the population SD across the ten fitted network levels.}
\label{tab:supp-profile}
\footnotesize
\begin{adjustbox}{max width=\textwidth}
\begin{tabular}{llllll}
\toprule
\multicolumn{6}{l}{\textit{A. Network-relative slopes [Bonferroni intervals across 40 network--state cells]}} \\
Network & REST & CARIT & FaceName & Visual--motor & Equal-state \\
\midrule
Visual1 & $-0.027$ [$-0.088$, $+0.035$] & $-0.047$ [$-0.116$, $+0.022$] & $-0.013$ [$-0.080$, $+0.054$] & $-0.081$ [$-0.168$, $+0.006$] & $-0.042$ \\
Visual2 & $+0.0004$ [$-0.035$, $+0.036$] & $-0.014$ [$-0.054$, $+0.025$] & $+0.023$ [$-0.017$, $+0.063$] & $-0.022$ [$-0.064$, $+0.021$] & $-0.003$ \\
Somatomotor & $-0.001$ [$-0.037$, $+0.034$] & $+0.004$ [$-0.034$, $+0.043$] & $+0.009$ [$-0.030$, $+0.049$] & $+0.038$ [$-0.005$, $+0.081$] & $+0.013$ \\
Cingulo-Opercular & $-0.003$ [$-0.037$, $+0.031$] & $+0.021$ [$-0.016$, $+0.058$] & $+0.002$ [$-0.035$, $+0.039$] & $+0.007$ [$-0.033$, $+0.047$] & $+0.007$ \\
Dorsal-attention & $-0.023$ [$-0.055$, $+0.009$] & $-0.018$ [$-0.055$, $+0.018$] & $-0.015$ [$-0.051$, $+0.021$] & $-0.010$ [$-0.046$, $+0.027$] & $-0.016$ \\
Language & $+0.020$ [$-0.032$, $+0.072$] & $+0.043$ [$-0.015$, $+0.101$] & $+0.033$ [$-0.022$, $+0.088$] & $+0.052$ [$-0.010$, $+0.114$] & $+0.037$ \\
Frontoparietal & $+0.073$$^{*}$ [$+0.030$, $+0.116$] & $+0.056$$^{*}$ [$+0.010$, $+0.102$] & $+0.040$ [$-0.007$, $+0.088$] & $+0.063$$^{*}$ [$+0.011$, $+0.116$] & $+0.058$ \\
Auditory & $-0.030$ [$-0.074$, $+0.013$] & $-0.055$$^{*}$ [$-0.104$, $-0.007$] & $-0.057$$^{*}$ [$-0.107$, $-0.007$] & $-0.038$ [$-0.095$, $+0.018$] & $-0.045$ \\
Default & $-0.017$ [$-0.057$, $+0.023$] & $+0.001$ [$-0.043$, $+0.044$] & $+0.002$ [$-0.044$, $+0.048$] & $-0.015$ [$-0.062$, $+0.033$] & $-0.007$ \\
Posterior-Multimodal & $+0.007$ [$-0.046$, $+0.060$] & $+0.011$ [$-0.049$, $+0.071$] & $-0.025$ [$-0.083$, $+0.033$] & $+0.005$ [$-0.058$, $+0.068$] & $-0.001$ \\
\midrule
\multicolumn{6}{l}{\textit{B. Fitted-profile geometry [95\% multiplier intervals]}} \\
State & Baseline level--slope $r$ & 5-year level--slope $r$ & SD ratio, 5 y/baseline & FP residual from trend & \\
\midrule
REST & $-0.725$ [$-0.847$, $-0.406$] & $-0.674$ [$-0.814$, $-0.317$] & $0.932$ [$0.901$, $0.966$] & $+0.032$ [$+0.013$, $+0.053$] &  \\
CARIT & $-0.874$ [$-0.933$, $-0.616$] & $-0.845$ [$-0.916$, $-0.541$] & $0.908$ [$0.878$, $0.942$] & $+0.008$ [$-0.016$, $+0.032$] &  \\
FaceName & $-0.707$ [$-0.829$, $-0.350$] & $-0.655$ [$-0.791$, $-0.264$] & $0.936$ [$0.904$, $0.973$] & $+0.025$ [$-0.002$, $+0.052$] &  \\
Visual--motor & $-0.219$ [$-0.532$, $+0.081$] & $-0.093$ [$-0.442$, $+0.228$] & $0.980$ [$0.940$, $1.023$] & $+0.050$ [$+0.018$, $+0.082$] &  \\
Equal-state average & $-0.725$ [$-0.852$, $-0.431$] & $-0.669$ [$-0.818$, $-0.326$] & $0.926$ [$0.896$, $0.960$] & $+0.024$ [$+0.002$, $+0.046$] &  \\
\bottomrule
\end{tabular}
\end{adjustbox}
\end{table}

Pairwise comparisons placed FP at the positive end of the reconfiguration without isolating it (Table~\ref{tab:supp-fp-comparisons}). Its four-state trajectory differed from 8 of the 9 other networks after max-Wald familywise adjustment; Posterior-Multimodal was not separated ($p=.218$). For the equal-state slope, FP exceeded 7 of 9 comparators with simultaneous intervals excluding zero; the Language and Posterior-Multimodal differences were not resolved, and the Somatomotor lower bound was $+0.0004$.

Uncentered network levels combine the common and relative components (Table~\ref{tab:supp-network-levels}). Under signal normalization, FP was the only network whose equal-state level increased with an interval excluding zero ($+0.070$ [$+0.005$, $+0.134$]). Estimated equal-state slopes of $H_k$ were negative in all ten networks. Under candidate total normalization, Visual2, Dorsal-attention and Auditory declined with intervals excluding zero, whereas the FP level slope was near zero ($+0.015$ [$-0.045$, $+0.075$]).

\begin{table}[!htbp]
\centering
\caption{FP-minus-comparator slope differences of uncentered signal-normalized network log levels (log units per 5 years). Familywise $p$ values are max-Wald adjusted across the nine four-state trajectory tests (4,999 pedigree-cluster multiplier draws; the smallest resolvable value is $1/(B+1)=.0002$). Equal-state intervals are max-$|t|$ simultaneous across the nine comparators and are given to four decimals because one bound lies close to zero.}
\label{tab:supp-fp-comparisons}
\footnotesize
\begin{adjustbox}{max width=\textwidth}
\begin{tabular}{lrrrrrcl}
\toprule
Comparator & REST & CARIT & FaceName & Visual--motor & Wald $\chi^2_4$ & Familywise $p$ & Equal-state difference [95\% CI] \\
\midrule
Visual1 & $+0.100$ & $+0.103$ & $+0.053$ & $+0.144$ & $35.24$ & $.0002$ & $+0.1002$ [$+0.0336$, $+0.1668$] \\
Visual2 & $+0.073$ & $+0.070$ & $+0.017$ & $+0.085$ & $36.60$ & $.0002$ & $+0.0613$ [$+0.0123$, $+0.1104$] \\
Somatomotor & $+0.074$ & $+0.052$ & $+0.031$ & $+0.025$ & $26.44$ & $.0002$ & $+0.0456$ [$+0.0004$, $+0.0908$] \\
Cingulo-Opercular & $+0.076$ & $+0.035$ & $+0.038$ & $+0.056$ & $25.62$ & $.0002$ & $+0.0514$ [$+0.0050$, $+0.0978$] \\
Dorsal-attention & $+0.096$ & $+0.074$ & $+0.055$ & $+0.073$ & $39.00$ & $.0002$ & $+0.0747$ [$+0.0313$, $+0.1181$] \\
Language & $+0.053$ & $+0.013$ & $+0.007$ & $+0.011$ & $16.44$ & $.0194$ & $+0.0212$ [$-0.0358$, $+0.0781$] \\
Auditory & $+0.103$ & $+0.112$ & $+0.097$ & $+0.102$ & $29.20$ & $.0002$ & $+0.1035$ [$+0.0486$, $+0.1584$] \\
Default & $+0.090$ & $+0.056$ & $+0.038$ & $+0.078$ & $30.89$ & $.0002$ & $+0.0653$ [$+0.0150$, $+0.1157$] \\
Posterior-Multimodal & $+0.066$ & $+0.045$ & $+0.065$ & $+0.058$ & $10.16$ & $.2180$ & $+0.0589$ [$-0.0022$, $+0.1199$] \\
\bottomrule
\end{tabular}
\end{adjustbox}
\end{table}

\begin{table}[!htbp]
\centering
\caption{Equal-state slopes of uncentered network log levels (log units per 5 years), from the profile model's weights, with Bonferroni intervals across the 30 entries; $^{*}$ interval excludes zero. $L^{\mathrm{TOT}}_k=L_k+H_k$. These slopes combine common and network-relative change; an interval that includes zero does not establish maintenance. They are not comparable to the separately weighted common-level slopes in Table~\ref{tab:supp-common}.}
\label{tab:supp-network-levels}
\footnotesize
\begin{adjustbox}{max width=\textwidth}
\begin{tabular}{llll}
\toprule
Network & Signal-normalized $L_k$ & $H_k$ & Candidate total-normalized $L_k^{\mathrm{TOT}}$ \\
\midrule
Visual1 & $-0.031$ [$-0.119$, $+0.057$] & $-0.032$ [$-0.123$, $+0.058$] & $-0.063$ [$-0.146$, $+0.021$] \\
Visual2 & $+0.008$ [$-0.051$, $+0.067$] & $-0.077$ [$-0.160$, $+0.006$] & $-0.069$$^{*}$ [$-0.135$, $-0.003$] \\
Somatomotor & $+0.024$ [$-0.033$, $+0.081$] & $-0.055$ [$-0.150$, $+0.039$] & $-0.032$ [$-0.109$, $+0.046$] \\
Cingulo-Opercular & $+0.018$ [$-0.031$, $+0.068$] & $-0.074$ [$-0.150$, $+0.001$] & $-0.056$ [$-0.120$, $+0.007$] \\
Dorsal-attention & $-0.005$ [$-0.061$, $+0.051$] & $-0.068$ [$-0.143$, $+0.007$] & $-0.073$$^{*}$ [$-0.128$, $-0.019$] \\
Language & $+0.048$ [$-0.025$, $+0.121$] & $-0.082$$^{*}$ [$-0.163$, $-0.001$] & $-0.034$ [$-0.112$, $+0.045$] \\
Frontoparietal & $+0.070$$^{*}$ [$+0.005$, $+0.134$] & $-0.054$ [$-0.133$, $+0.024$] & $+0.015$ [$-0.045$, $+0.075$] \\
Auditory & $-0.034$ [$-0.098$, $+0.029$] & $-0.101$ [$-0.209$, $+0.006$] & $-0.135$$^{*}$ [$-0.228$, $-0.042$] \\
Default & $+0.004$ [$-0.061$, $+0.069$] & $-0.067$ [$-0.149$, $+0.014$] & $-0.063$ [$-0.128$, $+0.001$] \\
Posterior-Multimodal & $+0.011$ [$-0.056$, $+0.078$] & $-0.081$ [$-0.165$, $+0.003$] & $-0.070$ [$-0.163$, $+0.022$] \\
\bottomrule
\end{tabular}
\end{adjustbox}
\end{table}

\section{Normalization comparison}
\label{sec:supp-normalization}

\begin{figure}[!htbp]
    \centering
    \includegraphics[width=\textwidth]{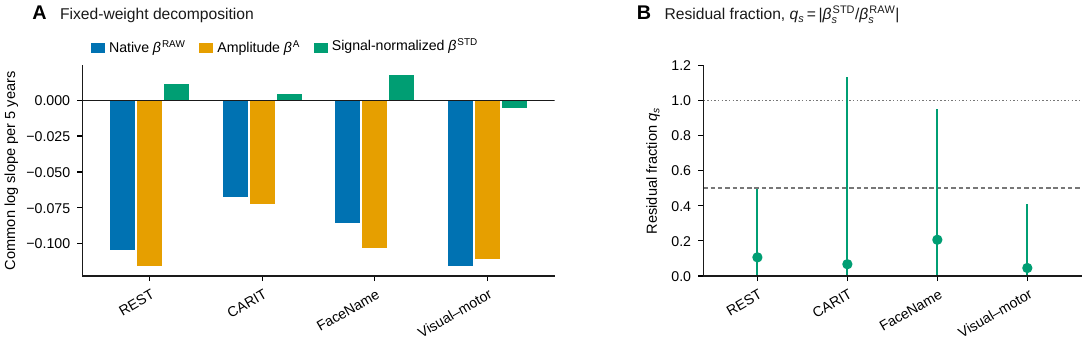}
    \caption{Fixed-weight decomposition of the native common slope. (A) Native slope $\beta^{\mathrm{RAW}}_s$ and its pair-amplitude ($\beta^{\mathrm{A}}_s$) and signal-normalized ($\beta^{\mathrm{STD}}_s$) components, all fitted with the native model's weights so that the two components sum to the native slope. (B) Residual fraction $q_s=|\beta^{\mathrm{STD}}_s/\beta^{\mathrm{RAW}}_s|$ with max-$|t|$ simultaneous 95\% intervals across states (4,999 pedigree-cluster multiplier draws). Dashed and dotted lines mark 0.5 and 1.}
    \label{fig:supp-residual-fraction}
\end{figure}

The fixed-weight decomposition of the native common slope (Table~\ref{tab:supp-common}A) is shown in Fig.~\ref{fig:supp-residual-fraction}.

\begin{figure}[!htbp]
    \centering
    \includegraphics[width=\textwidth]{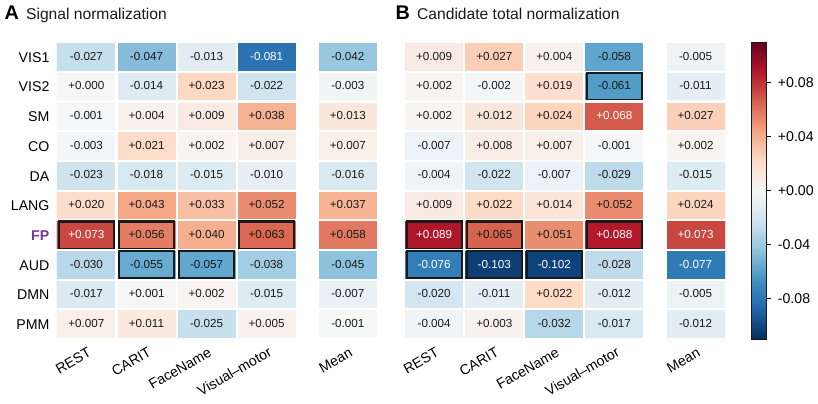}
    \caption{Network-relative profile under two amplitude normalizations. (A) Signal normalization (as in main-text Fig.~3A). (B) Candidate total normalization. Cells give state-specific slopes of $R_k$ in log units per 5 years; outlined cells have Bonferroni intervals across the 40 network--state cells that exclude zero. The Mean column is the equal-state average. Both panels use the same network order and color scale; numerical values and intervals for panel B are in Table~\ref{tab:supp-normalization}C.}
    \label{fig:supp-two-normalizations}
\end{figure}

Both profile tests remained significant under candidate total normalization (time$\times$network $\chi^2_9=35.98$, Holm $p=8.0\times10^{-5}$; time$\times$network$\times$state $\chi^2_{27}=155.62$, Holm $p=2.9\times10^{-19}$; Table~\ref{tab:supp-normalization}A), but individual cells changed (Fig.~\ref{fig:supp-two-normalizations}; Table~\ref{tab:supp-normalization}C). Seven cells excluded zero: FP in REST, CARIT and visual--motor; Auditory in REST, CARIT and FaceName; and Visual2 in visual--motor. The $H$ profile itself changed over time (time$\times$network Holm $p=.038$; time$\times$network$\times$state Holm $p=3.1\times10^{-8}$). FP-relative slopes were positive in all four states under candidate total normalization ($+0.056$ to $+0.099$; Table~\ref{tab:supp-normalization}B). The paired differences from signal normalization were $+0.010$ to $+0.028$ (four-state $\chi^2_4=5.79$, $p=.215$), with simultaneous intervals including zero in every state.

\begin{table}[!htbp]
\centering
\caption{Signal versus candidate total normalization of the network profile (log units per 5 years). All three representations use the signal-normalized profile model's GLS weights, so $L^{\mathrm{TOT}}=L^{\mathrm{STD}}+H$ holds exactly for every estimate and the paired difference equals the $H$ slope. Panel A: the six profile tests form one Holm family; FP-relative tests are nominal; FP was characterized after examining the profile in this cohort. Panel B: the paired $p$ is unadjusted; the four-state test of the paired differences is the $H$ row of Panel A. Panel C: $^{*}$ interval excludes zero; the equal-state column is the mean of the four point estimates. The corresponding signal-normalized estimates are in Table~\ref{tab:supp-profile}A.}
\label{tab:supp-normalization}
\footnotesize
\begin{adjustbox}{max width=\textwidth}
\begin{tabular}{llllll}
\toprule
\multicolumn{6}{l}{\textit{A. Joint tests (profile tests Holm-adjusted across six; FP tests nominal)}} \\
Representation & Time$\times$network & Time$\times$network$\times$state & FP-relative, four states & FP-relative, state heterogeneity & \\
\midrule
Signal-normalized & $\chi^2_{9}=43.90$, $p_{\mathrm{Holm}}=4.42\times10^{-6}$ & $\chi^2_{27}=112.00$, $p_{\mathrm{Holm}}=1.27\times10^{-11}$ & $\chi^2_{4}=37.01$, $p=1.79\times10^{-7}$ & $\chi^2_{3}=12.68$, $p=.0054$ &  \\
$H$ (total $-$ signal) & $\chi^2_{9}=17.73$, $p_{\mathrm{Holm}}=.0384$ & $\chi^2_{27}=90.82$, $p_{\mathrm{Holm}}=3.14\times10^{-8}$ & $\chi^2_{4}=5.79$, $p=.215$ & $\chi^2_{3}=3.52$, $p=.318$ &  \\
Candidate total-normalized & $\chi^2_{9}=35.98$, $p_{\mathrm{Holm}}=8.00\times10^{-5}$ & $\chi^2_{27}=155.62$, $p_{\mathrm{Holm}}=2.91\times10^{-19}$ & $\chi^2_{4}=34.43$, $p=6.09\times10^{-7}$ & $\chi^2_{3}=12.28$, $p=.0065$ &  \\
\midrule
\multicolumn{6}{l}{\textit{B. FP-relative slopes [simultaneous 95\% CI across four states]}} \\
State & Signal-normalized & Candidate total-normalized & Paired difference ($H$) & Paired $p$ & \\
\midrule
REST & $+0.081$ [$+0.044$, $+0.118$] & $+0.099$ [$+0.051$, $+0.147$] & $+0.018$ [$-0.012$, $+0.047$] & $.131$ &  \\
CARIT & $+0.062$ [$+0.023$, $+0.102$] & $+0.072$ [$+0.020$, $+0.125$] & $+0.010$ [$-0.025$, $+0.045$] & $.470$ &  \\
FaceName & $+0.045$ [$+0.004$, $+0.085$] & $+0.056$ [$+0.004$, $+0.108$] & $+0.012$ [$-0.021$, $+0.044$] & $.367$ &  \\
Visual--motor & $+0.070$ [$+0.025$, $+0.115$] & $+0.098$ [$+0.042$, $+0.154$] & $+0.028$ [$-0.006$, $+0.061$] & $.0381$ &  \\
\midrule
\multicolumn{6}{l}{\textit{C. Network-relative slopes under candidate total normalization [Bonferroni across 40 cells]}} \\
Network & REST & CARIT & FaceName & Visual--motor & Equal-state \\
\midrule
Visual1 & $+0.009$ [$-0.062$, $+0.079$] & $+0.027$ [$-0.050$, $+0.103$] & $+0.004$ [$-0.071$, $+0.079$] & $-0.058$ [$-0.149$, $+0.032$] & $-0.005$ \\
Visual2 & $+0.002$ [$-0.051$, $+0.054$] & $-0.002$ [$-0.058$, $+0.054$] & $+0.019$ [$-0.037$, $+0.074$] & $-0.061$$^{*}$ [$-0.119$, $-0.004$] & $-0.011$ \\
Somatomotor & $+0.002$ [$-0.061$, $+0.065$] & $+0.012$ [$-0.052$, $+0.076$] & $+0.024$ [$-0.041$, $+0.089$] & $+0.068$ [$-0.003$, $+0.138$] & $+0.027$ \\
Cingulo-Opercular & $-0.007$ [$-0.057$, $+0.044$] & $+0.008$ [$-0.046$, $+0.061$] & $+0.007$ [$-0.048$, $+0.062$] & $-0.001$ [$-0.058$, $+0.056$] & $+0.002$ \\
Dorsal-attention & $-0.004$ [$-0.051$, $+0.042$] & $-0.022$ [$-0.073$, $+0.030$] & $-0.007$ [$-0.057$, $+0.043$] & $-0.029$ [$-0.080$, $+0.023$] & $-0.015$ \\
Language & $+0.009$ [$-0.056$, $+0.074$] & $+0.022$ [$-0.049$, $+0.094$] & $+0.014$ [$-0.054$, $+0.083$] & $+0.052$ [$-0.023$, $+0.126$] & $+0.024$ \\
Frontoparietal & $+0.089$$^{*}$ [$+0.034$, $+0.145$] & $+0.065$$^{*}$ [$+0.004$, $+0.126$] & $+0.051$ [$-0.010$, $+0.111$] & $+0.088$$^{*}$ [$+0.023$, $+0.154$] & $+0.073$ \\
Auditory & $-0.076$$^{*}$ [$-0.150$, $-0.002$] & $-0.103$$^{*}$ [$-0.179$, $-0.028$] & $-0.102$$^{*}$ [$-0.176$, $-0.029$] & $-0.028$ [$-0.109$, $+0.052$] & $-0.077$ \\
Default & $-0.020$ [$-0.081$, $+0.041$] & $-0.011$ [$-0.079$, $+0.058$] & $+0.022$ [$-0.043$, $+0.086$] & $-0.012$ [$-0.084$, $+0.059$] & $-0.005$ \\
Posterior-Multimodal & $-0.004$ [$-0.074$, $+0.066$] & $+0.003$ [$-0.072$, $+0.079$] & $-0.032$ [$-0.105$, $+0.041$] & $-0.017$ [$-0.095$, $+0.060$] & $-0.012$ \\
\bottomrule
\end{tabular}
\end{adjustbox}
\end{table}

\section{Site and technical robustness}
\label{sec:supp-robustness}

\begin{figure}[!htbp]
    \centering
    \includegraphics[width=\textwidth]{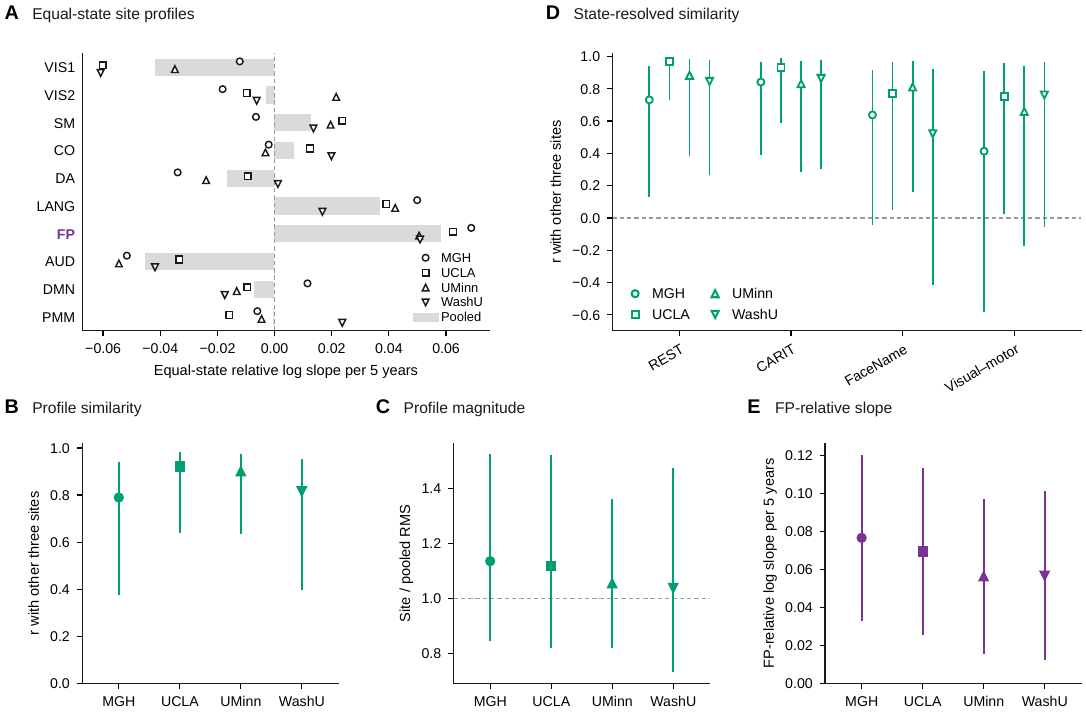}
    \caption{Site-specific profile-shape diagnostics (signal normalization). (A) Model-implied equal-state network-relative slopes at each site (symbols) and pooled (bars). (B) Correlation of each site profile with the mean profile of the other three sites. (C) Ratio of site to pooled profile root-mean-square magnitude; the dashed line marks one. (D) State-resolved correlations with the other three sites. (E) Equal-state FP-relative slope at each site. Intervals are simultaneous 95\% intervals within each family (4,999 pedigree-cluster multiplier draws).}
    \label{fig:supp-site-profile}
\end{figure}

Site-specific equal-state profiles were extracted as contrasts from the pooled joint profile model (Table~\ref{tab:supp-site}). Correlations between each site's profile and the mean profile of the other three sites ranged from .790 to .922, and all simultaneous intervals excluded zero. Among the six site pairs, the weakest correlation was MGH--WashU ($r=.638$ [$.067$, $.894$]). All profile-magnitude ratio intervals included one. The equal-state FP-relative slope was positive at every site ($+0.056$ to $+0.077$), with all four simultaneous intervals excluding zero. The pooled profile nevertheless varied across sites: the site$\times$time$\times$network and site$\times$time$\times$state$\times$network tests were significant under signal normalization ($\chi^2_{27}=56.13$, $p=8.3\times10^{-4}$; $\chi^2_{81}=205.95$, $p=7.5\times10^{-13}$) and under the other representations (Table~\ref{tab:supp-site}C). Site-specific profiles are displayed in Fig.~\ref{fig:supp-site-profile}, and site composition and state coverage are given in Table~\ref{tab:supp-site-composition}.

\begin{table}[!htbp]
\centering
\caption{Model-implied equal-state site profiles, extracted as contrasts from the pooled joint profile model (signal normalization). Correlation intervals are simultaneous within the ten-correlation family (Fisher-$z$ scale), RMS-ratio intervals within the four ratios (log scale), and FP-relative intervals within the four sites; all from 4,999 pedigree-cluster multiplier draws. Site-specific profile Wald tests (9 df) are reported as nominal descriptive tests. Panel C gives the site interaction tests of the pooled profile models; they are nominal and not part of the Holm family.}
\label{tab:supp-site}
\footnotesize
\begin{adjustbox}{max width=\textwidth}
\begin{tabular}{lcccc}
\toprule
\multicolumn{5}{l}{\textit{A. Each site}} \\
Site & Profile Wald $p$ & $r$ with other three & Site/pooled RMS & FP-relative slope \\
\midrule
MGH & $1.98\times10^{-6}$ & $0.790$ [$0.376$, $0.941$] & $1.135$ [$0.846$, $1.524$] & $+0.077$ [$+0.033$, $+0.120$] \\
UCLA & $4.63\times10^{-4}$ & $0.922$ [$0.639$, $0.985$] & $1.118$ [$0.821$, $1.522$] & $+0.069$ [$+0.026$, $+0.113$] \\
UMinn & $5.85\times10^{-6}$ & $0.902$ [$0.635$, $0.976$] & $1.055$ [$0.818$, $1.361$] & $+0.056$ [$+0.015$, $+0.097$] \\
WashU & $.0018$ & $0.816$ [$0.396$, $0.954$] & $1.038$ [$0.731$, $1.472$] & $+0.057$ [$+0.012$, $+0.101$] \\
\midrule
\multicolumn{5}{l}{\textit{B. Site pairs}} \\
Pair & $r$ [95\% CI] & & & \\
\midrule
MGH--UCLA & $0.779$ [$0.295$, $0.945$] & & &  \\
MGH--UMinn & $0.824$ [$0.484$, $0.948$] & & &  \\
MGH--WashU & $0.638$ [$0.067$, $0.894$] & & &  \\
UCLA--UMinn & $0.870$ [$0.506$, $0.971$] & & &  \\
UCLA--WashU & $0.879$ [$0.439$, $0.979$] & & &  \\
UMinn--WashU & $0.786$ [$0.294$, $0.949$] & & &  \\
\midrule
\multicolumn{5}{l}{\textit{C. Site heterogeneity of the ten-network profile (nominal tests)}} \\
Representation & Site$\times$time$\times$network & Site$\times$time$\times$state$\times$network & & \\
\midrule
Native & $\chi^2_{27}=44.84$, $p=.0169$ & $\chi^2_{81}=183.01$, $p=7.36\times10^{-10}$ & &  \\
Signal-normalized & $\chi^2_{27}=56.13$, $p=8.28\times10^{-4}$ & $\chi^2_{81}=205.95$, $p=7.55\times10^{-13}$ & &  \\
Candidate total-normalized & $\chi^2_{27}=65.24$, $p=5.21\times10^{-5}$ & $\chi^2_{81}=191.56$, $p=6.03\times10^{-11}$ & &  \\
\bottomrule
\end{tabular}
\end{adjustbox}
\end{table}

\begin{table}[!htbp]
\centering
\caption{Site composition and state coverage. Analyzed rows are session--state observations (8,727 in total). Follow-up is the median [interquartile range] in years. State entries are participants/analyzed rows. All participant identifiers had HCA prefixes; the event-level study label (HCA or AABC) was treated as visit-level provenance.}
\label{tab:supp-site-composition}
\footnotesize
\begin{adjustbox}{max width=\textwidth}
\begin{tabular}{lrrrccrcccc}
\toprule
Site & Participants & Sessions & Analyzed rows & Baseline age, mean [range] & Follow-up, y & Scanner changes & REST & CARIT & FaceName & Visual--motor \\
\midrule
MGH & 218 & 578 & 2,270 & 61.4 [36, 89] & 5.60 [3.62, 6.40] & 0 & 218/578 & 218/566 & 217/559 & 218/567 \\
UCLA & 193 & 464 & 1,813 & 58.9 [36, 89] & 5.14 [2.34, 6.29] & 8 & 193/464 & 193/454 & 193/449 & 192/446 \\
UMinn & 221 & 590 & 2,330 & 61.7 [36, 88] & 5.70 [4.02, 6.23] & 4 & 221/590 & 221/581 & 220/580 & 221/579 \\
WashU & 220 & 588 & 2,314 & 61.6 [36, 89] & 5.14 [3.39, 6.42] & 18 & 220/588 & 220/578 & 220/577 & 220/571 \\
\midrule
Total & 852 & 2,220 & 8,727 & & & 30 & & & &  \\
\bottomrule
\end{tabular}
\end{adjustbox}
\end{table}

\begin{figure}[!htbp]
    \centering
    \includegraphics[width=\textwidth]{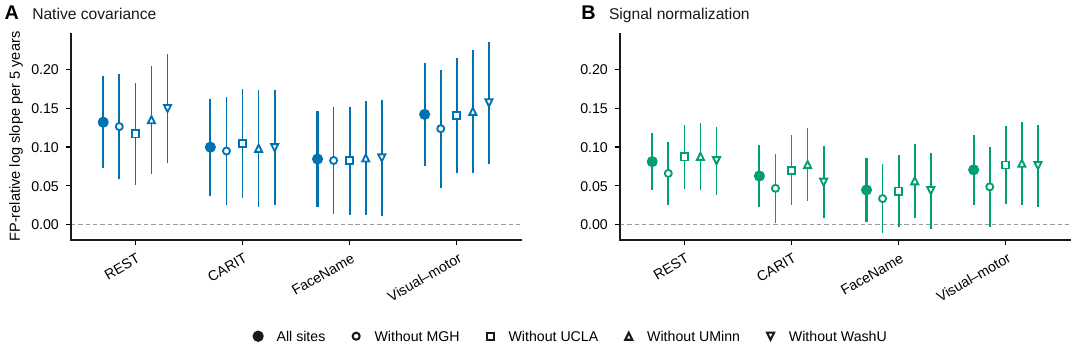}
    \caption{Leave-one-site-out FP-relative slopes. Filled symbols show the all-site estimate and open symbols the estimate after omitting each site; vertical lines are simultaneous 95\% intervals across the four states. (A) Native covariance, shown for comparison; native FP-relative slopes also include network differences in pair-amplitude change. (B) Signal normalization, the primary representation.}
    \label{fig:supp-loso}
\end{figure}

Omitting one site at a time left all 16 signal-normalized FP-relative point estimates positive (REST $+0.066$ to $+0.088$; CARIT $+0.047$ to $+0.077$; FaceName $+0.033$ to $+0.056$; visual--motor $+0.049$ to $+0.079$), with wider intervals after the loss of about a quarter of the sample (Fig.~\ref{fig:supp-loso}). The FP-relative model did not detect site$\times$time or site$\times$time$\times$state interactions under signal normalization ($p=.552$ and $p=.135$; Table~\ref{tab:supp-robustness}C).

\begin{figure}[!htbp]
    \centering
    \includegraphics[width=\textwidth]{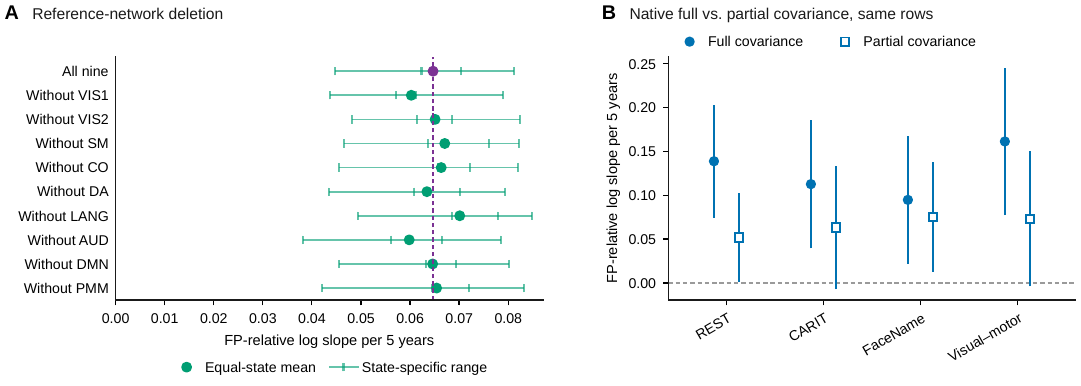}
    \caption{Reference-network and covariance sensitivity of the FP-relative slope. (A) Signal normalization: equal-state slope (points) and range of the four state-specific slopes (lines) after deleting each network from the nine-network reference; the purple point and dashed line mark the all-nine reference. Values come from the joint profile fit. (B) FP-relative slopes from native full covariance and native partial covariance fitted to the same 6,143 session--state rows, taken from the ten-network profile models ($\tfrac{10}{9}R_{\mathrm{FP}}$), with Bonferroni 95\% intervals across states.}
    \label{fig:supp-reference-partial}
\end{figure}

Deleting each network in turn from the nine-network reference gave equal-state FP-relative slopes of 0.0598--0.0701, compared with 0.0647 with all nine; the largest absolute change was 0.0054, and all four-state joint tests remained significant ($p\leq9.14\times10^{-7}$; Table~\ref{tab:supp-robustness}A; Fig.~\ref{fig:supp-reference-partial}A).

Partial covariance was available for 848 participants, 2,210 sessions and 6,143 session--state rows, and was compared with native full covariance on the same rows (Table~\ref{tab:supp-robustness}B; Fig.~\ref{fig:supp-reference-partial}B). FP-relative point estimates remained positive in all four states under partial covariance but were smaller on average (equal-state mean $+0.066$ versus $+0.127$); these differences are descriptive. In the direct FP-relative model, the partial-covariance four-state test was $\chi^2_4=9.79$, $p=.044$. The equal-state profile test was not detected under partial covariance (time$\times$network $p=.137$, versus $p=5.2\times10^{-5}$ for full covariance on the same rows), whereas state-dependent reconfiguration remained detected (time$\times$network$\times$state $p=.0029$).

\begin{table}[!htbp]
\centering
\caption{FP-relative robustness (log units per 5 years). Panel A: the nine-network reference recomputed after deleting each network; values come from the joint profile fit. Panel B: both columns are native (not amplitude-normalized) representations fitted to the same rows with available partial covariance; FP-relative slopes are $\tfrac{10}{9}R_{\mathrm{FP}}$ from the ten-network profile models, with Bonferroni intervals across the four states computed from the scaled standard errors. The difference between columns is descriptive; no paired test was performed. In the direct FP-minus-nine model, the partial-covariance four-state test was $\chi^2_4=9.79$, $p=.0440$. Panel C: nominal site interaction tests.}
\label{tab:supp-robustness}
\footnotesize
\begin{adjustbox}{max width=\textwidth}
\begin{tabular}{llll}
\toprule
\multicolumn{4}{l}{\textit{A. Reference-network deletion (equal-state FP-relative slope; joint four-state test)}} \\
Reference & Equal-state slope & Change from all nine & Joint $p$ \\
\midrule
All nine & $+0.0647$ & $+0.0000$ & $1.79\times10^{-7}$ \\
Without Visual1 & $+0.0603$ & $-0.0044$ & $9.14\times10^{-7}$ \\
Without Visual2 & $+0.0651$ & $+0.0004$ & $3.27\times10^{-7}$ \\
Without Somatomotor & $+0.0671$ & $+0.0024$ & $1.67\times10^{-7}$ \\
Without Cingulo-Opercular & $+0.0663$ & $+0.0017$ & $1.61\times10^{-7}$ \\
Without Dorsal-attention & $+0.0634$ & $-0.0013$ & $5.85\times10^{-7}$ \\
Without Language & $+0.0701$ & $+0.0054$ & $6.62\times10^{-8}$ \\
Without Auditory & $+0.0598$ & $-0.0049$ & $1.67\times10^{-7}$ \\
Without Default & $+0.0646$ & $-0.0001$ & $4.61\times10^{-7}$ \\
Without Posterior-Multimodal & $+0.0654$ & $+0.0007$ & $2.33\times10^{-8}$ \\
\midrule
\multicolumn{4}{l}{\textit{B. Native full versus native partial covariance on the same rows (profile-model FP-relative slopes)}} \\
State & Full covariance & Partial covariance & \\
\midrule
REST & $+0.139$ [$+0.074$, $+0.203$] & $+0.052$ [$+0.001$, $+0.103$] &  \\
CARIT & $+0.113$ [$+0.039$, $+0.186$] & $+0.063$ [$-0.007$, $+0.134$] &  \\
FaceName & $+0.095$ [$+0.022$, $+0.167$] & $+0.075$ [$+0.013$, $+0.137$] &  \\
Visual--motor & $+0.161$ [$+0.077$, $+0.245$] & $+0.073$ [$-0.004$, $+0.150$] &  \\
Equal-state mean (point estimate) & $+0.127$ & $+0.066$ &  \\
Profile time$\times$network & $\chi^2_{9}=35.35$, $p=5.17\times10^{-5}$ & $\chi^2_{9}=13.62$, $p=.137$ &  \\
Profile time$\times$network$\times$state & $\chi^2_{27}=112.13$, $p=2.42\times10^{-12}$ & $\chi^2_{27}=51.73$, $p=.0029$ &  \\
Participants/sessions/rows & \multicolumn{2}{c}{848/2,210/6,143 (both columns)} &  \\
\midrule
\multicolumn{4}{l}{\textit{C. Site interactions in the FP-relative model ($p$)}} \\
Representation & Site$\times$time & Site$\times$time$\times$state & \\
\midrule
Native & $.358$ ($\chi^2_{3}=3.23$) & $.0104$ ($\chi^2_{9}=21.55$) &  \\
Signal-normalized & $.552$ ($\chi^2_{3}=2.10$) & $.135$ ($\chi^2_{9}=13.66$) &  \\
Partial covariance & $.801$ ($\chi^2_{3}=1.00$) & $.516$ ($\chi^2_{9}=8.18$) &  \\
\bottomrule
\end{tabular}
\end{adjustbox}
\end{table}

\section{Structural sensitivity}
\label{sec:supp-structure}

\begin{figure}[!htbp]
    \centering
    \includegraphics[width=\textwidth]{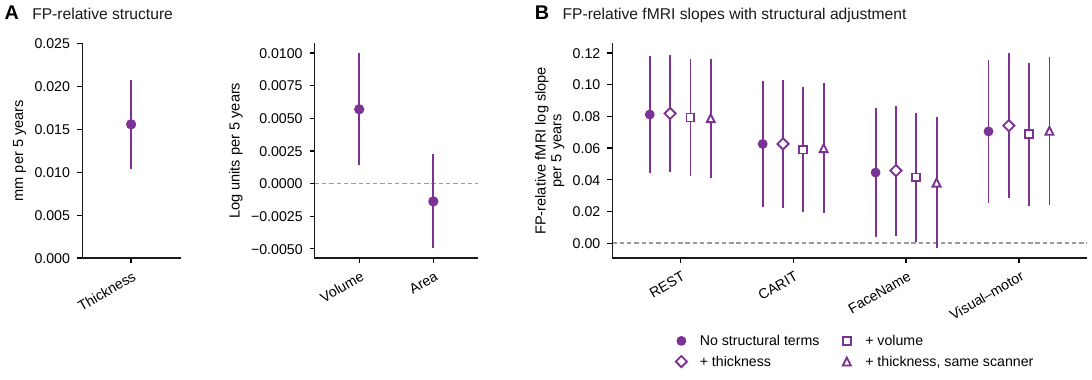}
    \caption{Structural sensitivity. (A) FP-relative slopes of cortical thickness (mm per 5 years), ICV-adjusted log volume and log surface area (log units per 5 years), with 95\% intervals. (B) Signal-normalized FP-relative fMRI slopes without structural terms, with within- and between-person thickness or volume terms, and with thickness terms in participants scanned on a single scanner; simultaneous 95\% intervals across states.}
    \label{fig:supp-structure}
\end{figure}

FP cortical thickness declined less than the nine-network reference ($+0.0156$ mm per 5 years [$0.0104$, $0.0207$], $p=2.93\times10^{-9}$), and FP-relative ICV-adjusted log volume showed a positive slope ($+0.00569$ [$0.00139$, $0.00998$], $p=.0094$); the relative surface-area slope was small ($-0.00137$, $p=.457$; Table~\ref{tab:supp-structure}A; Fig.~\ref{fig:supp-structure}A).

Adding within- and between-person thickness terms changed the norm of the four-state FP-relative fMRI slope vector by $+2.1$\%, and adding volume terms changed it by $-3.7$\%. Joint tests of the four adjusted-minus-unadjusted slope differences did not detect a change for either adjustment ($p=.719$ and $p=.206$; Table~\ref{tab:supp-structure}B; Fig.~\ref{fig:supp-structure}B). The thickness-adjusted slopes remained positive in every state among 822 participants scanned on a single scanner ($p=2.78\times10^{-7}$) and after omitting each site (Table~\ref{tab:supp-structure}C). Within-person thickness coefficients were not jointly different from zero ($p=.671$); the within-person volume coefficients had a joint $p=.0485$ without detectable state heterogeneity ($p=.261$; Table~\ref{tab:supp-structure}D).

\begin{table}[!htbp]
\centering
\caption{Structural sensitivity. Positive relative structural slopes indicate a more positive FP trajectory than the nine-network reference. Panel B: norm change is $100(\lVert\beta_{\mathrm{adj}}\rVert/\lVert\beta_{\mathrm{unadj}}\rVert-1)$ for the four-state FP-relative fMRI slope vectors on matched rows; the joint test is a 4-df Wald test that the four adjusted-minus-unadjusted slope differences are all zero, with their covariance computed from the joint pedigree-clustered influence functions of the two fits. Panel C gives participants/rows and the four-state joint test; $^{\dagger}$ simultaneous interval includes zero. Panel D: within- and between-person structural terms from the adjusted models.}
\label{tab:supp-structure}
\footnotesize
\begin{adjustbox}{max width=\textwidth}
\begin{tabular}{lllll}
\toprule
\multicolumn{5}{l}{\textit{A. FP-relative structural trajectories}} \\
Measure & Slope per 5 years & 95\% CI & $p$ & \\
\midrule
Cortical thickness (mm) & $+0.0156$ & [$+0.0104$, $+0.0207$] & $2.93\times10^{-9}$ &  \\
ICV-adjusted log volume & $+0.00569$ & [$+0.00139$, $+0.00998$] & $.0094$ &  \\
Log surface area & $-0.00137$ & [$-0.00496$, $+0.00223$] & $.457$ &  \\
\midrule
\multicolumn{5}{l}{\textit{B. Change in the four-state FP-relative fMRI slope vector}} \\
Adjustment & Norm change & Joint $\chi^2_4$ & $p$ & \\
\midrule
Between-person thickness & $-0.2$\% & $0.54$ & $.969$ &  \\
Within- and between-person thickness & $+2.1$\% & $2.09$ & $.719$ &  \\
Within- and between-person volume & $-3.7$\% & $5.91$ & $.206$ &  \\
\midrule
\multicolumn{5}{l}{\textit{C. Thickness-adjusted FP-relative fMRI slopes in restricted samples}} \\
Sample & REST & CARIT & FaceName & Visual--motor \\
\midrule
All sites (852/8,727; $p=1.75\times10^{-7}$) & $+0.0818$ & $+0.0626$ & $+0.0458$ & $+0.0741$ \\
Same scanner (822/8,410; $p=2.78\times10^{-7}$) & $+0.0788$ & $+0.0599$ & $+0.0382$$^{\dagger}$ & $+0.0709$ \\
Without MGH (634/6,457; $p=5.22\times10^{-4}$) & $+0.0657$ & $+0.0458$ & $+0.0335$$^{\dagger}$ & $+0.0518$$^{\dagger}$ \\
Without UCLA (659/6,914; $p=4.33\times10^{-7}$) & $+0.0875$ & $+0.0710$ & $+0.0436$$^{\dagger}$ & $+0.0785$ \\
Without UMinn (631/6,397; $p=6.39\times10^{-6}$) & $+0.0886$ & $+0.0767$ & $+0.0577$ & $+0.0824$ \\
Without WashU (632/6,413; $p=5.41\times10^{-6}$) & $+0.0848$ & $+0.0556$ & $+0.0469$$^{\dagger}$ & $+0.0827$ \\
\midrule
\multicolumn{5}{l}{\textit{D. Structural coefficients in the FP-relative fMRI model (per 1 SD)}} \\
Component & REST & CARIT & FaceName & Visual--motor \\
\midrule
Thickness, between; joint $p=.0970$ & $-0.0060$ & $+0.0047$ & $-0.0055$ & $+0.0159$ \\
Thickness, within; joint $p=.671$ & $-0.0004$ & $+0.0012$ & $-0.0017$ & $-0.0077$ \\
Volume, between; joint $p=.229$ & $+0.0039$ & $+0.0047$ & $-0.0063$ & $+0.0133$ \\
Volume, within; joint $p=.0485$ & $+0.0036$ & $+0.0133$ & $+0.0092$ & $+0.0047$ \\
\bottomrule
\end{tabular}
\end{adjustbox}
\end{table}

\section{Baseline-age moderation}
\label{sec:supp-age}

Baseline age did not clearly modify the principal longitudinal effects over the observed age range (Table~\ref{tab:supp-age}). The equal-state baseline-age$\times$time$\times$network test for the signal-normalized profile gave $p=.0527$; the FP-relative moderation test gave $p=.748$.

\begin{table}[!htbp]
\centering
\caption{Joint tests of baseline-age moderation of longitudinal change.}
\label{tab:supp-age}
\small
\begin{tabular}{p{0.62\textwidth}rrr}
\toprule
Analysis & $\chi^2$ & df & $p$ \\
\midrule
Signal-normalized ten-network profile: age$\times$time$\times$network (equal-state) & $16.75$ & 9 & $.0527$ \\
Signal-normalized FP-relative trajectory: age$\times$time$\times$state & $1.93$ & 4 & $.748$ \\
Native common covariance: age$\times$time$\times$state & $7.44$ & 4 & $.114$ \\
Signal-normalized common level: age$\times$time$\times$state & $4.27$ & 4 & $.371$ \\
\bottomrule
\end{tabular}
\end{table}

\section{Cognition}
\label{sec:supp-cognition}

\begin{figure}[!htbp]
    \centering
    \includegraphics[width=\textwidth]{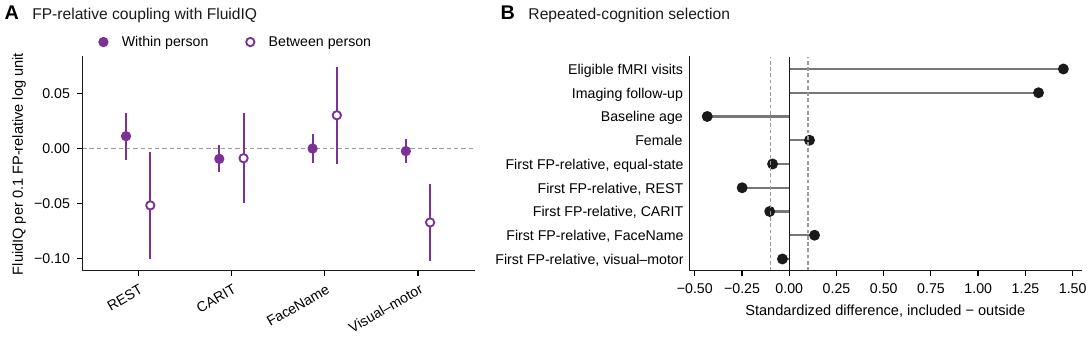}
    \caption{Cognition coefficients and repeated-cognition selection. (A) Within-person (filled) and between-person (open) FluidIQ coefficients per 0.1 FP-relative log unit, with simultaneous 95\% intervals across states. (B) Standardized differences between the 746 participants in the repeated-cognition analysis and those outside it; dashed lines mark $\pm0.1$.}
    \label{fig:supp-cognition}
\end{figure}

The repeated-cognition analysis included 746 participants (87.6\% of the imaging cohort) with at least two visits containing both four-state FP-relative fMRI and FluidIQ. The four within-person coefficients were not jointly different from zero ($\chi^2_4=5.63$, $p=.228$) and did not differ detectably across states ($\chi^2_3=4.08$, $p=.253$). Standardized within-person coefficients ranged from $-0.057$ to $+0.040$, and all simultaneous intervals included zero (Table~\ref{tab:supp-cognition}A; Fig.~\ref{fig:supp-cognition}A). Between-person coefficients were jointly different from zero ($\chi^2_4=36.58$, $p=2.20\times10^{-7}$), with intervals excluding zero in REST and visual--motor.

Participants in the repeated-cognition subset differed from the others mainly in longitudinal follow-up and age (Table~\ref{tab:supp-cognition}B; Fig.~\ref{fig:supp-cognition}B): they had more eligible fMRI visits (mean 2.61 versus 1.69), longer imaging follow-up (4.97 versus 2.23 years) and younger baseline age (60.1 versus 67.0 years), whereas the first equal-state FP-relative value differed little (standardized difference $-0.09$). FluidIQ variance among the 746 participants was 0.632 times that among the 87 participants with a single paired visit (bootstrap 95\% interval 0.481--0.883).

\begin{table}[!htbp]
\centering
\caption{Cognition coupling and repeated-cognition selection. Within-person standardized coefficients use the observed within-person SDs of FluidIQ and the FP-relative measure; they are regression coefficients, not correlations. Per-0.1-log coefficients give the FluidIQ difference associated with a 0.1-unit FP-relative log-covariance difference. Panel B compares the 746 included participants with those outside the repeated-cognition subset (continuous variables $n=102$; categorical $n=106$).}
\label{tab:supp-cognition}
\footnotesize
\begin{adjustbox}{max width=\textwidth}
\begin{tabular}{llll}
\toprule
\multicolumn{4}{l}{\textit{A. FluidIQ coefficients [simultaneous 95\% CI across four states]}} \\
State & Within, standardized & Within, per 0.1 log & Between, per 0.1 log \\
\midrule
REST & $+0.040$ [$-0.036$, $+0.115$] & $+0.011$ [$-0.010$, $+0.033$] & $-0.052$ [$-0.100$, $-0.003$] \\
CARIT & $-0.057$ [$-0.132$, $+0.018$] & $-0.009$ [$-0.022$, $+0.003$] & $-0.009$ [$-0.050$, $+0.032$] \\
FaceName & $-0.0004$ [$-0.068$, $+0.068$] & $-0.0001$ [$-0.013$, $+0.013$] & $+0.030$ [$-0.014$, $+0.074$] \\
Visual--motor & $-0.017$ [$-0.090$, $+0.057$] & $-0.002$ [$-0.013$, $+0.008$] & $-0.067$ [$-0.103$, $-0.032$] \\
Joint tests & \multicolumn{2}{l}{within $\chi^2_4=5.63$, $p=.228$; equality $\chi^2_3=4.08$, $p=.253$} & $\chi^2_4=36.58$, $p=2.20\times10^{-7}$ \\
\midrule
\multicolumn{4}{l}{\textit{B. Repeated-cognition selection}} \\
Variable & Included ($n=746$) & Outside subset & Standardized difference \\
\midrule
Eligible fMRI visits, mean & $2.61$ & $1.69$ & $+1.45$ \\
Imaging follow-up, years & $4.97$ & $2.23$ & $+1.32$ \\
Baseline age, years & $60.1$ & $67.0$ & $-0.43$ \\
Female, proportion & $0.582$ & $0.528$ & $+0.11$ \\
First equal-state FP-relative value & $-0.536$ & $-0.521$ & $-0.09$ \\
FluidIQ variance ratio, $\geq2$ vs.\ 1 paired visit & \multicolumn{3}{l}{$0.632$ [$0.481$, $0.883$]} \\
\bottomrule
\end{tabular}
\end{adjustbox}
\end{table}

